\documentclass[10pt,journal,compsoc, twoside]{IEEEtran}

\usepackage{multirow}

\usepackage{setspace}

\usepackage{subfigure}  

\usepackage{amsmath,amssymb,amsfonts}
\usepackage{textcomp}

\usepackage[linesnumbered, ruled, vlined]{algorithm2e}

\usepackage{booktabs}

\renewcommand\labelenumi{(\roman{enumi})}
\renewcommand\theenumi\labelenumi

\makeatletter
\def\hlinew#1{%
	\noalign{\ifnum0=`}\fi\hrule \@height #1 \futurelet
	\reserved@a\@xhline}

\ifCLASSOPTIONcompsoc
  \usepackage[nocompress]{cite}
\else
  \usepackage{cite}
\fi

\ifCLASSINFOpdf
   \usepackage[pdftex]{graphicx}
\else
   \usepackage[dvips]{graphicx}
\fi

\begin{document}

\title{Measurement and Analysis of the Bitcoin\\ Networks: A View from Mining Pools}

\author{Canhui~Wang,~\IEEEmembership{Graduate Student Member,~IEEE,}
        Xiaowen~Chu,~\IEEEmembership{Senior Member,~IEEE,}
        and~Qin~Yang,~\IEEEmembership{Senior~Member,~IEEE}
        
\IEEEcompsocitemizethanks{
	
	\IEEEcompsocthanksitem C. Wang is with Department of Computer Science, Hong Kong Baptist University, Kowloon Tong, Kowloon, Hong Kong, China. E-mail: chwang@comp.hkbu.edu.hk.

	\IEEEcompsocthanksitem X. Chu is with Department of Computer Science, Hong Kong Baptist University, Kowloon Tong, Kowloon, Hong Kong, China. E-mail: chxw@comp.hkbu.edu.hk.

	\IEEEcompsocthanksitem Q. Yang is with Department of Computer Science and Technology, Harbin Institute of Technology Shenzhen Graduate School, Shenzhen, China. E-mail: csyqin@hitsz.edu.cn.}

\thanks{Corresponding author: Xiaowen Chu. Our blockchain homepage is available at: https://blockchain.comp.hkbu.edu.hk/index.php.}

}

\IEEEtitleabstractindextext{%

\begin{abstract}
	
Bitcoin network, with the market value of \$68 billion as of January 2019, has received much attention from both industry and the academy. Mining pools, the main components of the Bitcoin network, dominate the computing resources and play essential roles in network security and performance aspects. Although many existing measurements of the Bitcoin network are available, little is known about the details of mining pool behaviors (e.g., empty blocks, mining revenue and transaction collection strategies) and their effects on the Bitcoin end users (e.g., transaction fees, transaction delay and transaction acceptance rate). This paper aims to fill this gap with a systematic study of mining pools. We traced over 1.56 hundred thousand blocks (including about 257 million historical transactions) from February 2016 to January 2019 and collected over 120.25 million unconfirmed transactions from March 2018 to January 2019. Then we conducted a board range of measurements on the pool evolutions, labeled transactions (blocks) as well as real-time network traffics, and discovered new interesting observations and features. Specifically, our measurements show the following. 1) A few mining pools entities continuously control most of the computing resources of the Bitcoin network. 2) Mining pools are caught in a prisoner's dilemma where mining pools compete to increase their computing resources even though the unit profit of the computing resource decreases. 3) Mining pools are stuck in a Malthusian trap where there is a stage at which the Bitcoin incentives are inadequate for feeding the exponential growth of the computing resources. 4) The market price and transaction fees are not sensitive to the event of halving block rewards. 5) The block interval of empty blocks is significantly lower than the block interval of non-empty blocks. 6) \textit{Feerate} plays a dominating role in transaction collection strategy for the top mining pools. Our measurements and analysis help to understand and improve the Bitcoin network.

\end{abstract}

\begin{IEEEkeywords}
Bitcoin Network, Mining Pools, Malthusian Trap, Incentive Mechanism
\end{IEEEkeywords}}

\maketitle
\IEEEdisplaynontitleabstractindextext
\IEEEpeerreviewmaketitle

\ifCLASSOPTIONcompsoc
\IEEEraisesectionheading{\section{Introduction}\label{sec:introduction}}
\else
\section{Introduction}
\label{sec:introduction}
\fi

\IEEEPARstart{B}{itcoin} \cite{nakamoto2008bitcoin} is a decentralized peer to peer (P2P) cryptocurrency that was first proposed by Satoshi Nakamoto in 2008. Without resorting to any trusted third party, Bitcoin adapts a cryptographic proof mechanism that enables anonymous peers to complete transactions through the P2P network. Blockchain is the core mechanism of the Bitcoin system. It not only records historical transactions from Bitcoin clients, but also prevents the Bitcoin network from double spending attacks \cite{karame2015misbehavior}. The Bitcoin network participants, who maintain and update the ongoing chain of blocks, are called miners. These miners compete in a mining race driven by an incentive mechanism \cite{lewenberg2015bitcoin, schrijvers2016incentive}, where the one who first solves the Bitcoin cryptographic puzzle \cite{giechaskiel2016bitcoin} has the right to collect unconfirmed transactions into a new block, append the new block to the main chain, i.e., the longest chain of blocks, and gain some BTCs \cite{BTC} as a mining reward.

With the popularity of Bitcoin, the cryptographic puzzles are getting increasingly difficult that ordinary miners cannot successfully solve it in ten minutes even though mining rewards are rewarding (i.e., block rewards of 12.5 BTCs per block, pricing at \$4,020 per BTC as of January 2019) \cite{Blockchaininfo}. As a result, a new mining approach named pooled mining \cite{SlushPool} becomes popular among miners since year 2010. The approach requires miner to trust the pool operator. Miners request mining tasks from the operator and return the computing results through the Stratum protocol\cite{eyal2015miner}. Once new blocks are mined, the mining rewards are redistributed to miners through the reward functions \cite{schrijvers2016incentive}. This approach enables miners to stabilize mining revenues by sharing market risks with other miners \cite{lewenberg2015bitcoin}.

Recently, mining pools have become the major computing resources of the Bitcoin network, and attract researchers to work on it for three main reasons. First, mining pools are of great importance to both security and performance of the Bitcoin network \cite{velner2017smart, miller2015nonoutsourceable, conti2018survey}. Different from small cryptocurrency systems such as Zcash \cite{Zcash} and Bitcoin Gold \cite{Bitcoingold} that suffered from 51\% attacks, the Bitcoin network has huge computing power such that the 51\% attacks require a huge amount of computing power which is expensive \cite{hassani2018big, almukaynizi2018finding} and therefore economically infeasible. Second, the competition among mining pools is intense and many existing works\cite{johnson2014game, eyal2015miner, altman2018mining} studied these mining behaviors in the game theory. Third, mining pool's behavior significantly affects the Bitcoin end users since most of the users' transaction data is processed by the mining pool.

Thus, given the importance of Bitcoin mining pools, we conduct a systematic study of the Bitcoin network from a perspective of mining pools, and provides the detailed analysis of the collected data set. In particular, we focus on the top four mining pools, i.e., AntPool \cite{AntPool}, F2Pool \cite{F2Pool}, ViaBTC \cite{ViaBTC} and BTC.com \cite{BTCcom}. Our main contributions to the literature are summarized as follows.

\begin{enumerate}
	
	\item We performed large scale measurements on historical transactions and blocks, covering over 1.56 hundred thousand blocks (including 257 million historical transactions) from February 25, 2016, to January 3, 2019. Also, we developed a python tool for processing and collecting unconfirmed transactions at Bitcoin Full Node's Mempool, and collected over 120.25 million unconfirmed transaction from March 6, 2018, to January 3, 2019. We made both the python tool and the collected data set publicly available to the research community \cite{github2}.
	
	\item We conducted a detailed analysis of the collected data set from a perspective of mining pools, including computing power distribution, mining revenue, transaction delay and transaction collection strategy. We discovered new observations and features of the Bitcoin network. For example, we found that mining pools are caught in a prisoner's dilemma where mining pools compete to increase their computing resources even though the unit profit of the computing resource decreases. And mining pools are stuck in a Malthusian trap where there is a stage at which the Bitcoin incentives are inadequate for feeding the exponential growth of the computing resources. Moreover, we dissected mining pool's behaviors and found that \textit{feerate} plays a dominating role in transaction collection strategy for the top mining pools. The acceptance rate of the unconfirmed transactions could be around 90 percent on average if its \textit{feerate} ranks are in top $X$, where $X$ is the number of transactions of the next block. On the converse, the acceptance rate of the unconfirmed transactions could be around less than 5 percent if its \textit{feerate} ranks are out of top $2\times X$. Our measurements and analysis contribute to better understand the Bitcoin network.

\end{enumerate}

The remainder of this paper is organized as follows. Section 2 provides the background of Bitcoin mining pools, including Bitcoin related terminology, incentive mechanism and pool mining principles. Section 3 illustrates our measurement setup and describes the collected data set. Section 4 presents a detailed analysis of historical transactions at the Bitcoin blockchain, including computing power evolution, computing power distribution, mining revenue and empty block. Section 5 analyzes the unconfirmed transactions at Bitcoin Full Node's Mmepool, including the transaction delay, transaction acceptance rate and transaction collection strategy. Section 6 surveys related works on the Bitcoin network measurement. Finally, section 7 concludes the paper.

\section{Background of Bitcoin Mining Pools}

\subsection{Terminology}

The terminology used in Bitcoin mining pools is not well standardized. For the purpose of clarity, we listed some terms of mining pools and define them as follows.

\vspace{2pt}

\textit{Definition 2.1 (Mining Pool).} The mining pool is the pooling of computing power by coordinating miners via a specific network protocol. SlushPool, the first mining pool, was introduced when mining difficulty was getting so hard that small miners could even take years to solve the mining difficulty \cite{eyal2018majority}.

\vspace{2pt}

\textit{Definition 2.2 (Hash Rate).} The hash rate is the speed of solving the Bitcoin cryptographic puzzles in the units of hashes per second. Following the Blockchain website \cite{Blockchaininfo}, the hash rate of the Bitcoin network was estimated to be around 43.29 Ehash/s as of January 3, 2019. 

\vspace{2pt}

\textit{Definition 2.3 (Mining Difficulty).} The mining difficulty is the difficulty of solving the Bitcoin cryptographic puzzles. To achieve a relatively stable block generation speed, Bitcoin adapts a term named mining difficulty that is adjusted every 2016 blocks ensuring that the Bitcoin network was calculating $\frac{2^{48}}{2^{16}-1}$ hashes in ten minutes on average. In other words, given a mining difficulty $D$, the expected hash rate $H$ for the Bitcoin network to generate a new block is $H=\frac{D\times 2^{48}}{(2^{16}-1)\times 10 \;mins}$. 

\vspace{2pt}

\textit{Definition 2.4 (Block Interval).} The block interval is the interval between two blocks. According to the Bitcoin protocol, Bitcoin uses a 10 minute average block interval. That means new blocks are generated every ten minutes \cite{nakamoto2008bitcoin}.

\vspace{2pt}

\textit{Definition 2.5 (Mining Reward).} The mining reward consists of two parts: block reward and transaction fees, where block reward refers to the first transaction in a block (i.e., the coinbase transaction of a block) and transaction fees refer to the slight difference between the total amount of inputs and the total outputs of every non-coinbase transaction included in a block. The mining reward encourages miners to faithfully follow the Bitcoin protocol, from which the security of the Bitcoin network is derived \cite{nakamoto2008bitcoin}.

\vspace{2pt}

\textit{Definition 2.6 (Empty Block).} The empty block is not empty. An empty block is a block with only one transaction, i.e., the coinbase transaction which allocates the block rewards to the block creator.

\vspace{2pt}

\textit{Definition 2.7 (Mempool).} The Mempool refers to a collection of Bitcoin unconfirmed transactions. Each full node in the Bitcoin network holds unconfirmed transactions at local Mempool, and remove them once miners successfully append them to the Bitcoin blockchain.

\vspace{2pt}

\textit{Definition 2.8 (Transaction Collection Strategy).} The transaction collection strategy refers to the mining pool's strategy for collecting unconfirmed transactions into a new block. Many factors \cite{li2018transaction, croman2016scaling} affect it such as waiting time, transaction size and transaction fees. We will further discuss the transaction collection strategy in section 5.2.

\subsection{Incentive Mechanism}

An incentive mechanism \cite{sompolinsky2018bitcoin, eyal2018majority} is used to incent miners to work only on valid blocks so that invalid ones will be rejected, and eventually do not exist in the Bitcoin blockchain. Also, it attracts miners to continuously join and support the Bitcoin network. Specifically, the miner who successfully creates a new block is granted a certain amount of mining rewards. As mentioned above, mining rewards are consisted of a fixed amount of block rewards and the transaction fees of a block. Note that block rewards start with a subsidy of 50 BTCs. But the subsidy of block rewards is designed to be continuously halved after every 210,000 blocks (around every 4 years) and will eventually reach zero BTC when all 21 million BTCs are minted. Till that time, mining rewards will be only from transaction fees. We will further discuss block rewards transition in section 4.3.2.

\subsection{Pooled Mining Principles}

Pooled mining principles normally consist of two parts: share agreement and reward functions. The share agreement \cite{schrijvers2016incentive, eyal2018majority} means that all members of a mining pool work together to mine a new block and share mining revenues when any one of them successfully mines a new block. This allows each member of the mining pool to receive relatively stable mining rewards. The reward function \cite{schrijvers2016incentive} refers to the approach of dividing mining rewards by the pool operator. After completing the mining task, miners report their solutions to the pool operator. If the Bitcoin network accepts the solution, the operator collects the mining rewards and divide it among miners following a specific reward function such as PPS and PPLNS \cite{zhu2018survey} that both parties agreed with on beforehand.

\section{Measurement Setup}

\subsection{The Measurement Framework}

\begin{figure}[b]
	\centering
	\includegraphics[scale=0.94]{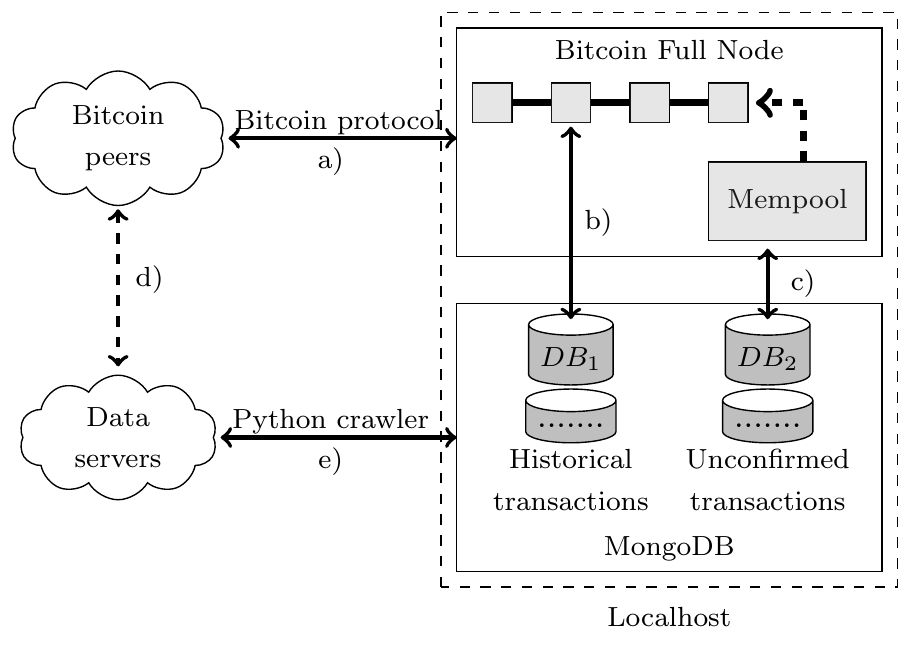}
	\caption{The Measurement Framework}
	\label{fig:universe}
\end{figure}

Fig. 1 shows the measurement framework. In step a$)$, we ran several Bitcoin Full Node clients to synchronize real time network traffics with other nodes in the Bitcoin network through the Bitcoin protocol. We then identified two types of data: historical transactions data and unconfirmed transactions data, where historical transactions data means transactions at the Bitcoin blockchain; and unconfirmed transactions data means unconfirmed transactions at Mempool. In step b$)$ and c$)$, we set up two MongoDB databases, denoted by $DB_1$ and $DB_2$, to store historical transactions data and unconfirmed transactions data, respectively. In step d$)$, most of the confirmed blocks with corresponding mining pool labels can be obtained from data servers such as the Blockchain.Info website and BTC.com. Finally, in step e$)$, we developed a python crawler to collect mining pool labels from BTC.com to identify the block creator of each confirmed block for further analysis.

\begin{table}[t]
	\centering
	\caption{Block Processing Results}
	\scalebox{0.8}{
		\begin{tabular}{|l|l|l|l|}
			\hline
			Category      & \# of Blocks & Proportion & Time Span (mm/dd/yy)    \\ \hline
			Known pools   & 153,842      & 98.18\%    & 02/25/2016 - 01/03/2019 \\ \hline
			Unknown pools & 2,853        & 1.82\%     & 02/25/2016 - 01/03/2019 \\ \hline
		\end{tabular}
	}
	\vspace{0pt}\\
	\setlength{\parindent}{-4.8em} {\scriptsize * Total number blocks from 02/25/2016 to 01/03/2019 is 156,695. }
	
\end{table}

\begin{table}[t]
	{\centering
		\caption{An Overview of Top 25 Bitcoin Mining Pools}
		\scalebox{0.82}{
			\begin{tabular}{|l|l|l|l|}
				\hline
				Mining Pool & Alias       & \# of Blocks & Time Span (mm/dd/yy) \\ \hline
				AntPool$^\star$     & N/A         & 27,026        & 02/25/2016 - 01/03/2019  \\ \hline
				F2Pool$^\star$      & Discus Fish & 19,282        & 02/25/2016 - 01/03/2019  \\ \hline
				BTC.com$^\star$     & Block Trail & 17,488        & 09/05/2016 - 01/03/2019  \\ \hline
				ViaBTC$^\star$      & N/A         & 12,100        & 06/05/2016 - 01/03/2019  \\ \hline
				SlushPool   & Bitcoin.cz  & 12,002        & 02/25/2016 - 01/03/2019  \\ \hline
				BTC.TOP     & N/A         & 11,256        & 12/11/2016 - 01/03/2019  \\ \hline
				BTCC        & BTCC China  & 10,586        & 02/25/2016 - 09/25/2018  \\ \hline
				BitFury     & N/A         & 8,754         & 02/25/2016 - 01/02/2019  \\ \hline
				BW.COM$^\dagger$      & LKETC       & 7,315         & 02/25/2016 - 04/03/2018  \\ \hline
				Bixin       & N/A         & 5,048         & 06/12/2016 - 01/02/2019  \\ \hline
				BitClub     & N/A         & 4,563         & 02/25/2016 - 01/03/2019  \\ \hline
				unknown     & N/A         & 2,853         & 03/11/2016 - 01/02/2019  \\ \hline
				GBMiners$^\dagger$    & N/A         & 2,093         & 08/30/2016 - 04/15/2018  \\ \hline
				1Hash$^\dagger$       & N/A         & 1,895         & 03/04/2016 - 12/07/2017  \\ \hline
				KanoPool    & N/A         & 1,813         & 02/25/2016 - 12/27/2018  \\ \hline
				Bitcoin.com & N/A         & 1,813         & 09/21/2016 - 01/02/2019  \\ \hline
				Poolin      & N/A         & 1,535          & 07/02/2018 - 01/03/2019  \\ \hline
				DPOOL       & N/A         & 1,199          & 03/31/2018 - 01/03/2019  \\ \hline
				KnCMiner$^\dagger$    & N/A         & 909          & 02/25/2016 - 09/12/2016  \\ \hline
				BTPOOL$^\dagger$      & N/A         & 703          & 06/15/2017 - 12/01/2017  \\ \hline
				58COIN      & N/A         & 694          & 11/05/2017 - 01/01/2019  \\ \hline
				Telco 214$^\dagger$   & Telcominer  & 653          & 02/27/2016 - 08/23/2017  \\ \hline
				WAYI.CN     & N/A         & 602          & 04/03/2018 - 01/02/2019  \\ \hline
				Huobi.pool     & N/A         & 578          & 04/12/2018 - 01/03/2019  \\ \hline
				BWPool      & N/A         & 517          & 04/03/2018 - 11/09/2018  \\ \hline
				
			\end{tabular}
		}
	}
	\vspace{4pt}
	\begin{spacing}{0.7}
		{\scriptsize * Total number blocks from 02/25/2016 to 01/03/2019 is 156695.}\\
		{\scriptsize * N/A indicates the mining pool does not provide the corresponding data .}\\
		{\scriptsize * The symbol $^\star$ indicates the top four major mining pools.}\\
		{\scriptsize * The symbol $^\dagger$ indicates the pool has no signs of activity for half a year.}\\
		{\scriptsize * Time span represents first observation time and last observation time from 02/25/2016 to 01/03/2019.}
	\end{spacing}
\end{table}

During the progress of data processing, for those blocks whose mining pool labels are available, we grouped them into known pools, meaning that the block creators are publicly known. Conversely, for those blocks whose mining pool labels are unavailable, we grouped them into unknown pools, meaning that the block creators are anonymous. Specifically, the data processing results are shown in TABLE 1. As we can see that 98.18 percent of block creators are available and only 1.49 percent of block creators are unavailable. The time span began on February 25, 2016, for the following two reasons. First, the computing power of the Bitcoin network increased dramatically and many well known mining pools such as BTC.com, ViaBTC and BTC.TOP started to appear after year 2016. Second, the Bitcoin's network computing power and market price \cite{Blockchaininfo} increased dramatically over the period, meaning that Bitcoin started to be popular among the public.

TABLE 2 shows an overview of top 25 Bitcoin mining pools from February 2016 to January 2019. According to BTC.com, there are many mining pools in the Bitcoin network. We crawled blocks of mining pools and traced related information such as alias, the number of blocks and time span. In the following sections, we will focus on analysis of top four minging pools, i.e., AntPool, F2Pool, BTC.com and ViaBTC.

\subsection{Challenges of Data Collection}

Our data sets are grouped into two types: historical transaction data and unconfirmed transaction data. The first type of data is stored in the local Bitcoin blockchain. The second type of data is obtained from real time network traffics disseminated across the Bitcoin network. Algorithm 1 shows the unconfirmed transaction timestamp algorithm. To solve the problem of inconsistent system clocks of different nodes in the Bitcoin network, we use the local system clock as the standard clock. We ran the Bitcoin Core 0.14.2 with the initial configuration of '\textit{{\small txindex=1}}'. Every two seconds, we performed a full scan of unconfirmed transactions at local Mempool using the '\textit{{\small bitcoin-cli getrawtransaction}}' command, and updated the corresponding observation time list. Once an unconfirmed transaction is successfully collected by miners, it will be removed from the Mempool. Thus, we locally maintained an observant timestamp list for each verified but not confirmed transaction at Mempool. Algorithm 2 shows the block processing algorithm. We obtained the block creators of 156,695 confirmed blocks from February 2016 to January 2019 via interacting with the data server BTC.com. Similarly, to solve the problem of inconsistent system clocks of block creators, we collected the timestamp that BTC.com first observed a block, and used this timestamp to calculate the block interval.

\begin{algorithm}[t]
	{\footnotesize \SetKwInOut{Input}{Input}\SetKwInOut{Output}{Output}
		\Input{Each unconfirmed transaction (\textit{transaction id}) at Mempool.}
		\Output{Each unconfirmed transaction (\textit{transaction id}) at Mempool, timestamped with the corresponding observation time list.}
		\BlankLine
		
		\While{number of unconfirmed transactions at Mempool $>$ 0}{
			
			request the Bitcoin \textit{Full Node} for each unconfirmed transaction (\textit{transaction id}) every two seconds; 
			
			\eIf{\rm the unconfirmed transaction (\textit{transaction id}) has not yet been removed from Mempool}{
				
				update the observation time list for the unconfirmed transaction (\textit{transaction id});
				
			}{
				continue;
			}	
	}}
	
	\caption{{\small Unconfirmed Transaction Timestamp}}
\end{algorithm}

\begin{algorithm}[t]
	
	{\footnotesize \SetKwInOut{Input}{Input}\SetKwInOut{Output}{Output}
		\Input{Each confirmed block (\textit{block id}); The data server \textit{BTC.com}.}
		\Output{Each confirmed block (\textit{block id}), labeled with corresponding mining pool label.}
		\BlankLine
		
		\While{number of confirmed blocks $>$ 0}{
			request the data server for each confirmed block (\textit{block id}).
			
			\eIf{\rm successfully get the mining pool label for the confirmed block (\textit{block id})}{
				update the mining pool label for the confirmed block (\textit{block id});
				
			}{
				label the confirmed block (\textit{block id}) as 'unknown';
				
			}
			
	}}
	\caption{{\small Block Processing Algorithm}}
\end{algorithm}

Three main challenges during the progress of data collection and data processing. The first challenge is the problem of inconsistent system clocks of transactions (blocks) creators as mentioned above, since nodes in the Bitcoin network could have a different machine system clock value. Our solution to the first challenge is to use a consistent machine system clock as the standard system clock. For unconfirmed transactions' data, we used the local system clock as the standard system clock, to calculate consistent transaction delays for different transactions. For confirmed blocks' data, we used the BTC.com system clock as the standard system clock, to calculate consistent block intervals for different blocks. The second challenge is data collection frequency. It means that we need to control the frequency of accessing local Bitcoin Full Node Mempool when maintaining the timestamp list for each unconfirmed transaction in the Mempool. For example, a higher frequency means a higher accuracy though a heavier cost of the data acquisition. Algorithm 1 shows our solution to the second challenge. To achieve a balance, we queried data every two seconds that is sufficient to satisfy the needs of maintaining the timestamp list for each unconfirmed transaction in the Mempool. The third problem is data loss. Various reasons such as network congestion and temporarily losing connections with the remote servers could cause the data loss problem. Algorithm 2 shows our solution to the third challenge. Every time we failed to get responses about block creator of a block; we will repeat requesting the remote server of BTC.com five times. If it still failed, we will label the block with a label of the unknown.

\section{Data Analysis on Historical Transactions}

We are interested in mining pool behavior characteristics, such as the evolution of mining pools, the computing power distribution, mining revenue and block size. To understand these characteristics, we will analyze labeled historical blocks and transactions in this section.

\subsection{Distribution of Computing Power}

\begin{figure}[b]
	\begin{center}
		\includegraphics[trim = 10mm 1mm 1mm 8mm, clip=true, width=9.9cm,scale=0.12]{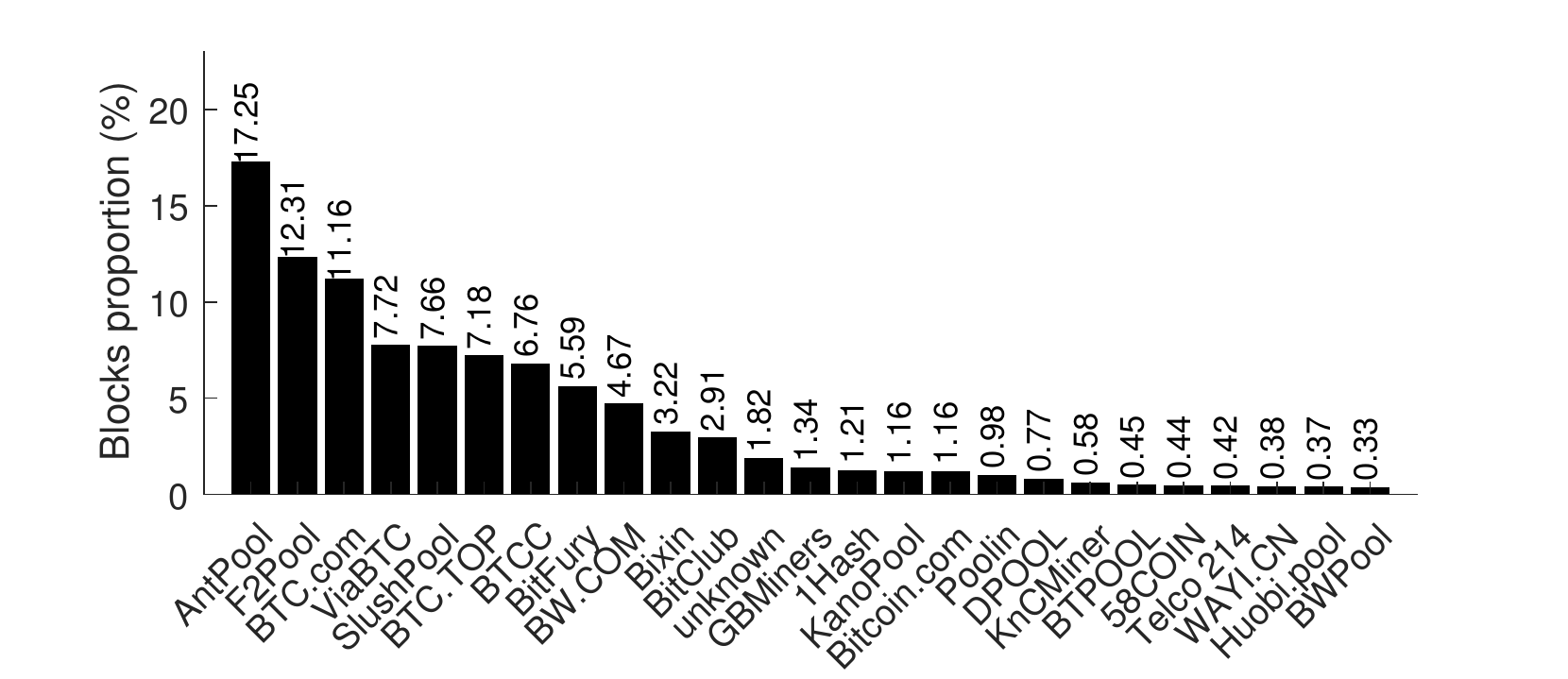}
	\end{center}
	\caption{An Estimation of Blocks Created by Mining Pools from Feb 25, 2016 to Jan 03, 2019}
\end{figure}

\begin{figure*}[htbp]
	\centering
	
	\subfigure[AntPool]{%
		\includegraphics[width=0.23\textwidth]{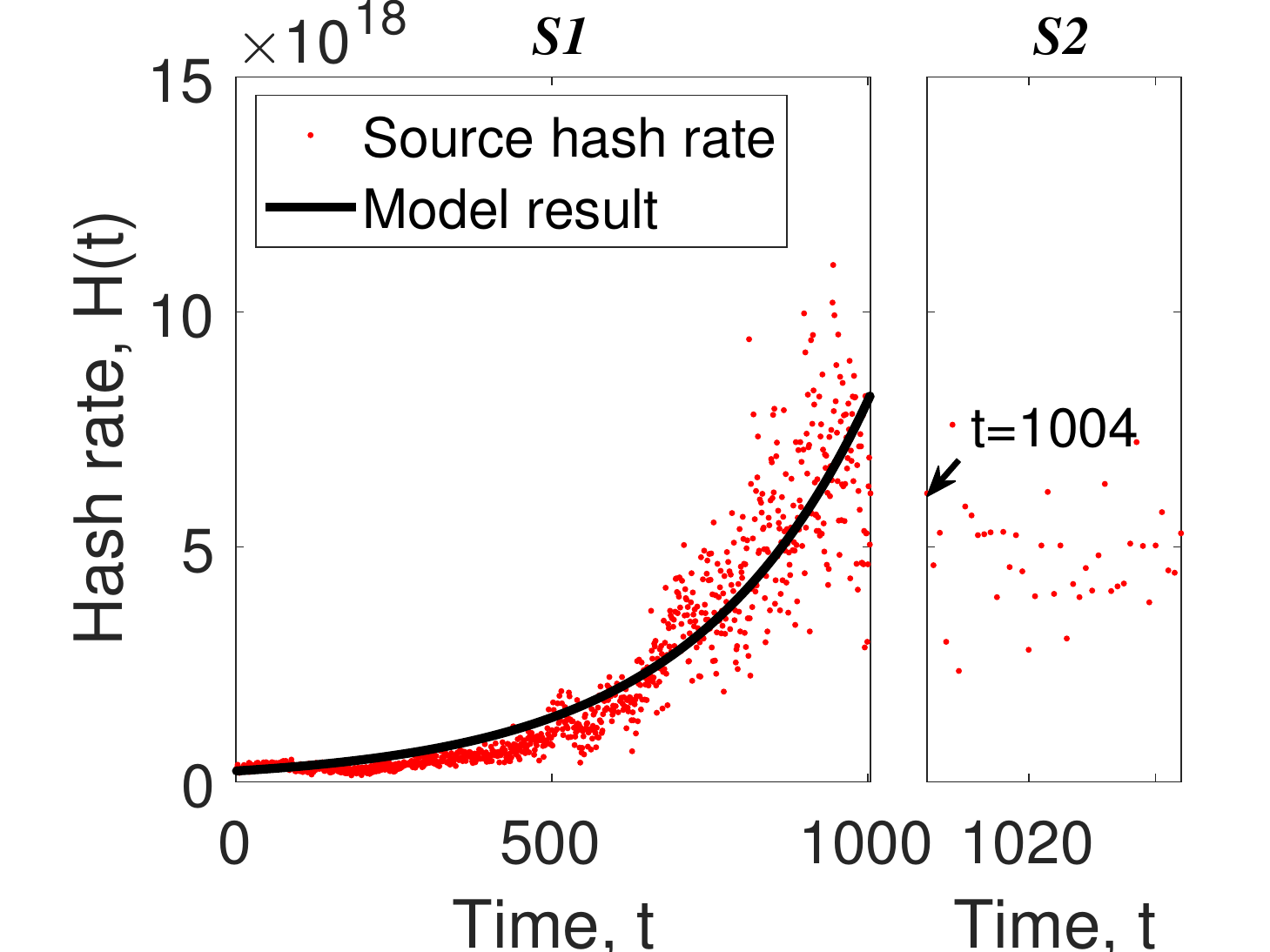}%
	}\hspace{0.2cm}
	\subfigure[F2Pool]{%
		\includegraphics[width=0.23\textwidth]{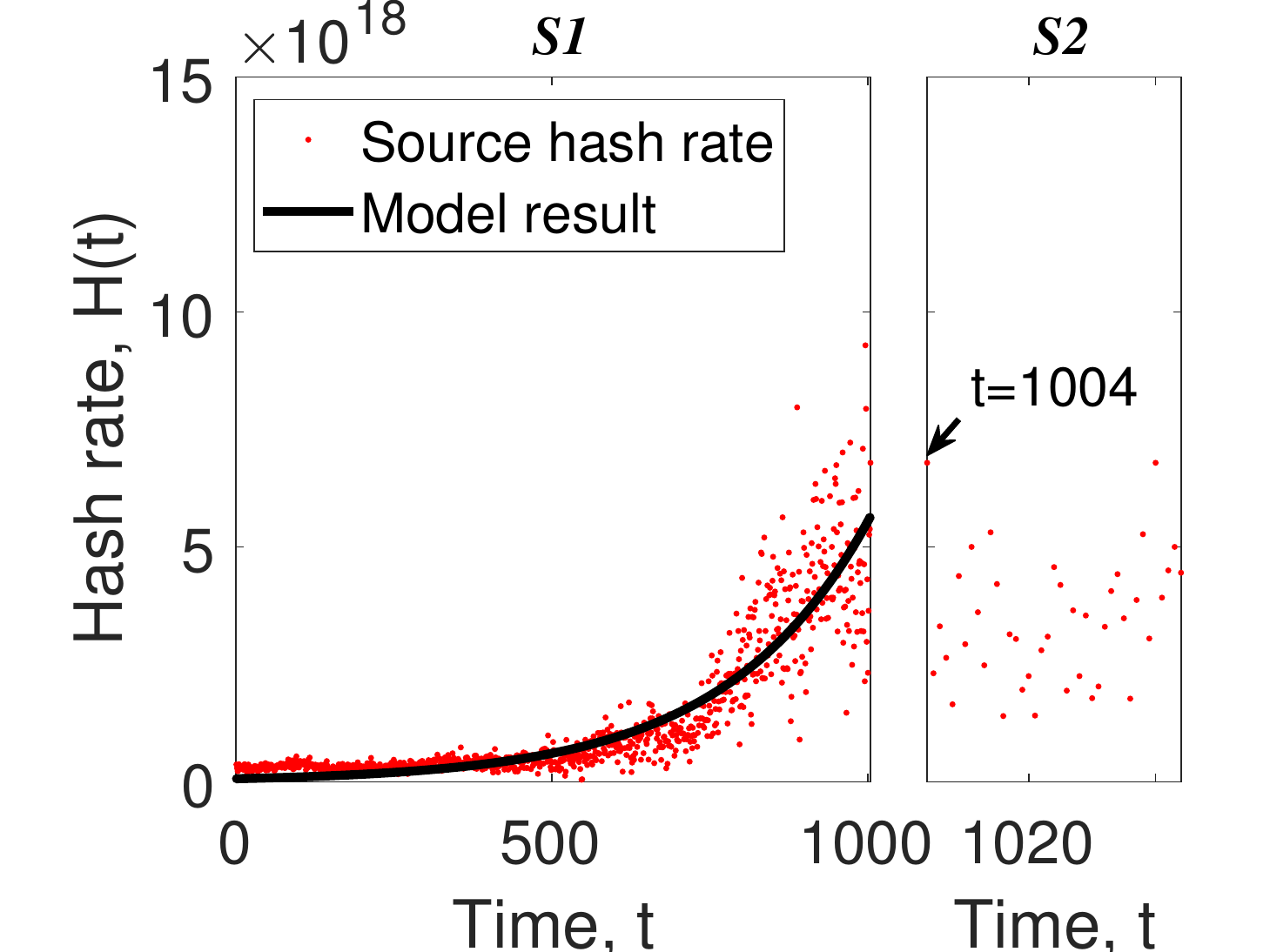}%
	}\hspace{0.2cm}
	\subfigure[ViaBTC]{%
		\includegraphics[width=0.23\textwidth]{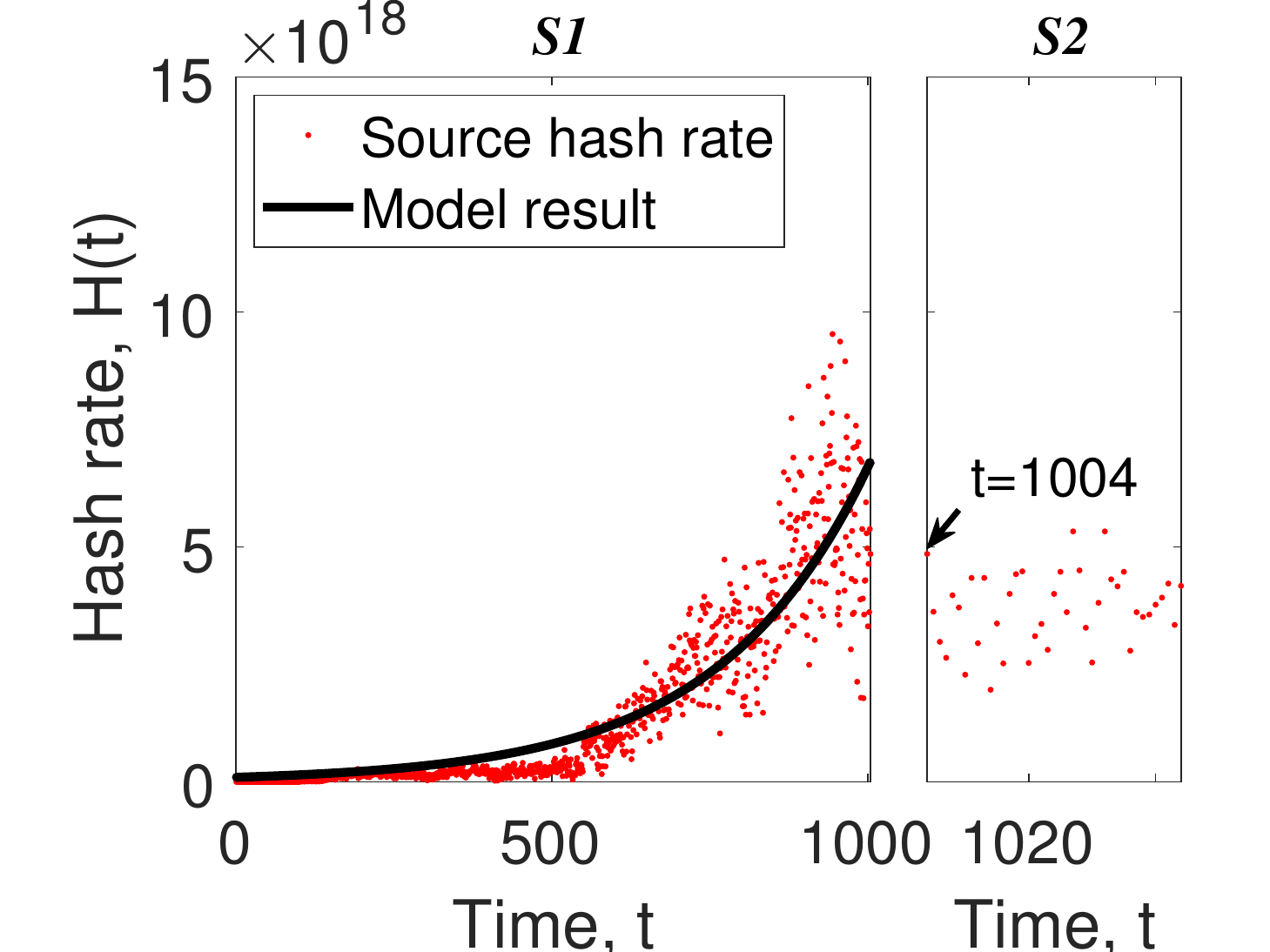}%
	}\hspace{0.2cm}
	\subfigure[BTC.com]{%
		\includegraphics[width=0.23\textwidth]{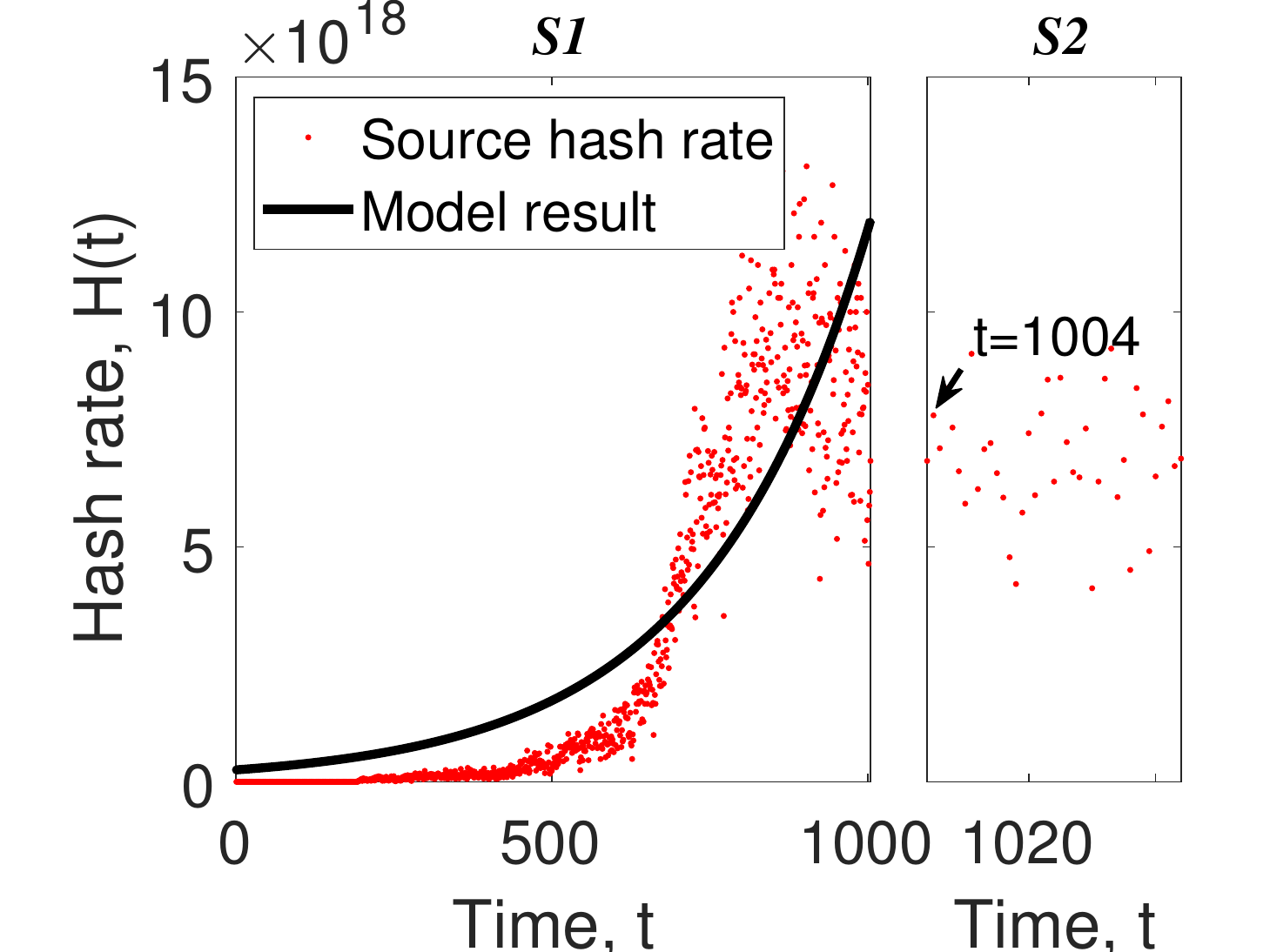}%
	}
	
	\caption{Daily Hash Rate of Top Mining Pools from Feb 25, 2016 to Jan 03, 2019}	
\end{figure*}

\begin{figure*}[t]
	\centering
	
	\subfigure[AntPool]{%
		\includegraphics[width=0.23\textwidth]{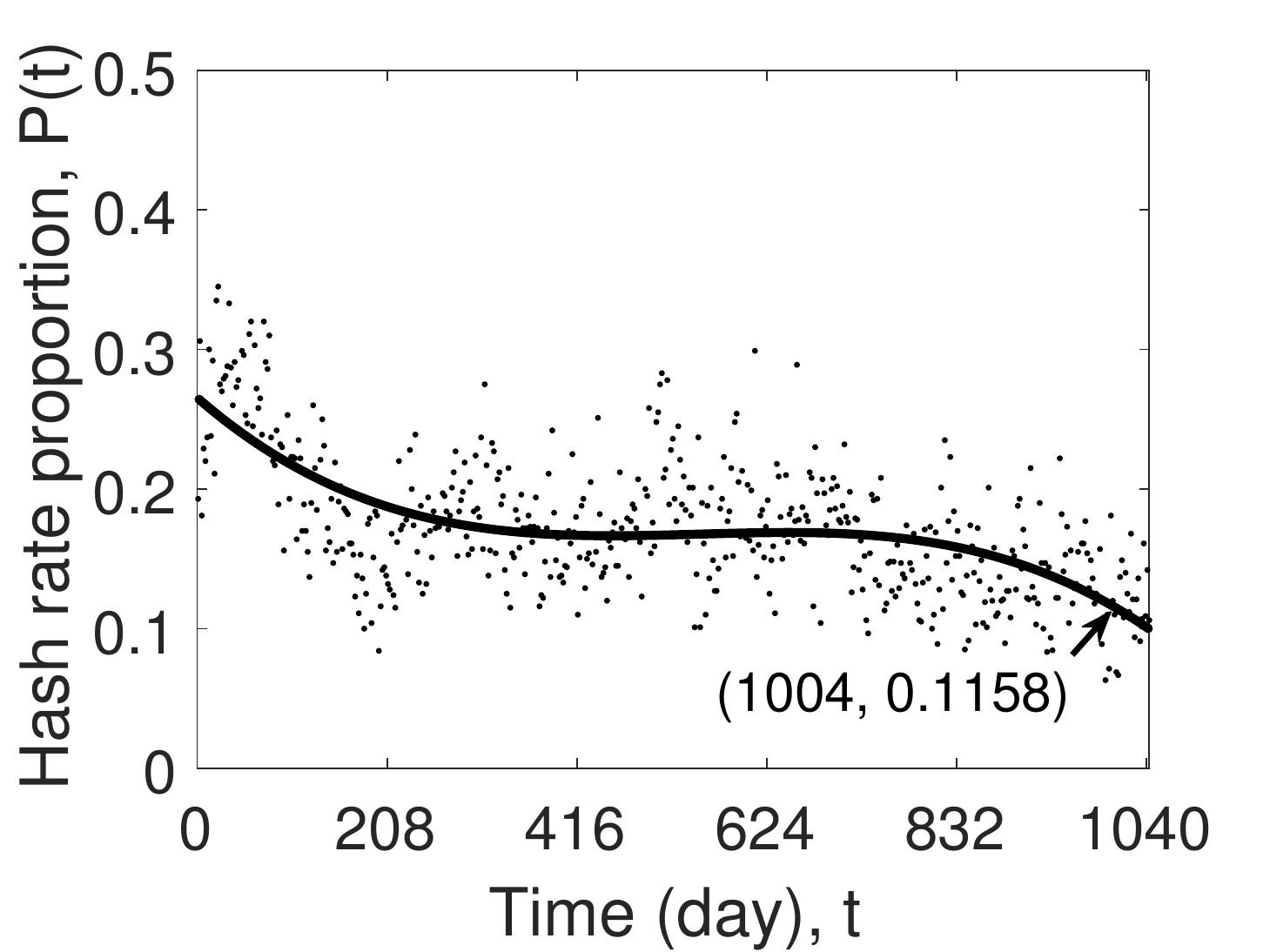}%
	}\hspace{0.2cm}
	\subfigure[F2Pool]{%
		\includegraphics[width=0.23\textwidth]{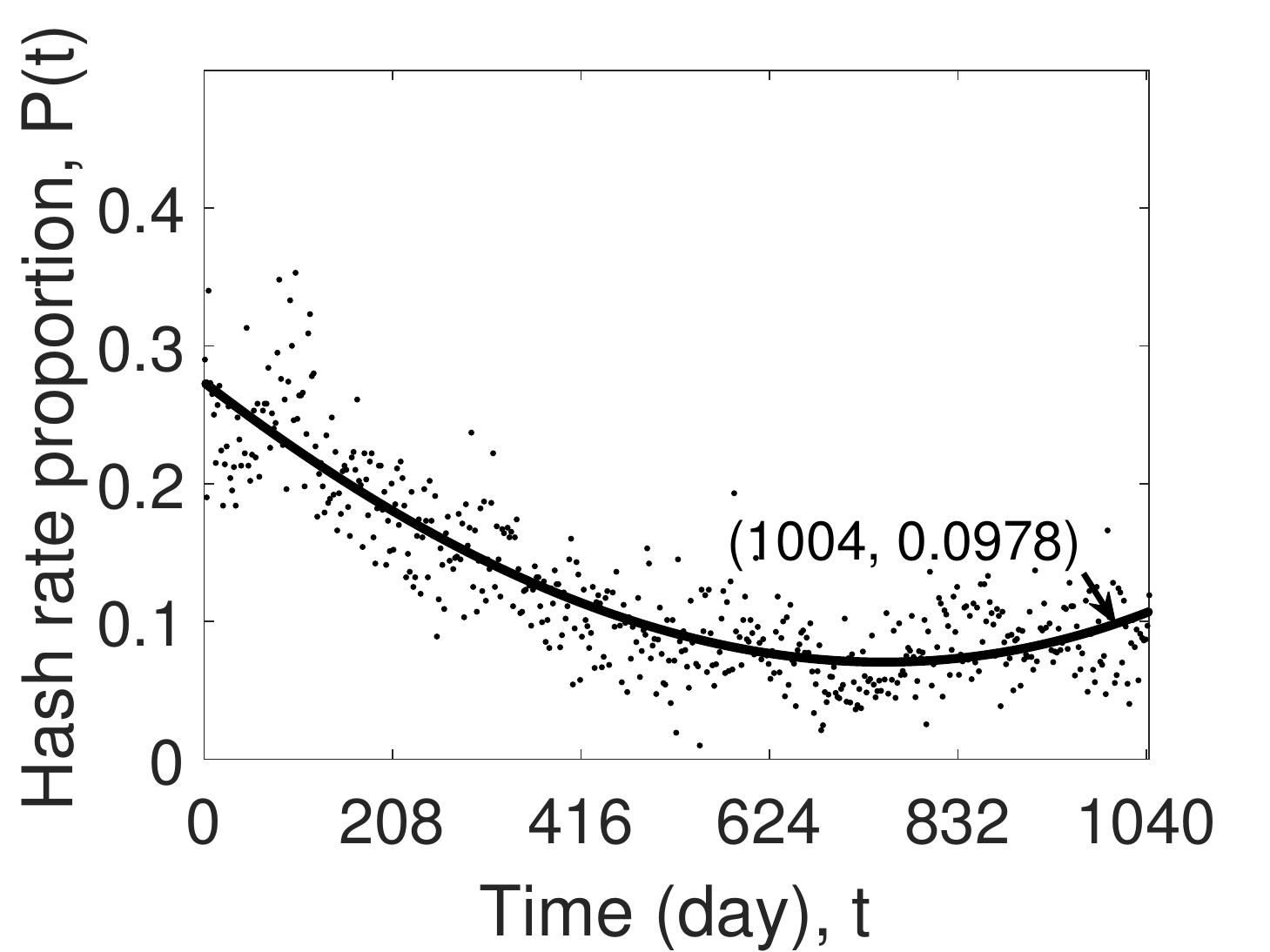}%
	}\hspace{0.2cm}
	\subfigure[ViaBTC]{%
		\includegraphics[width=0.23\textwidth]{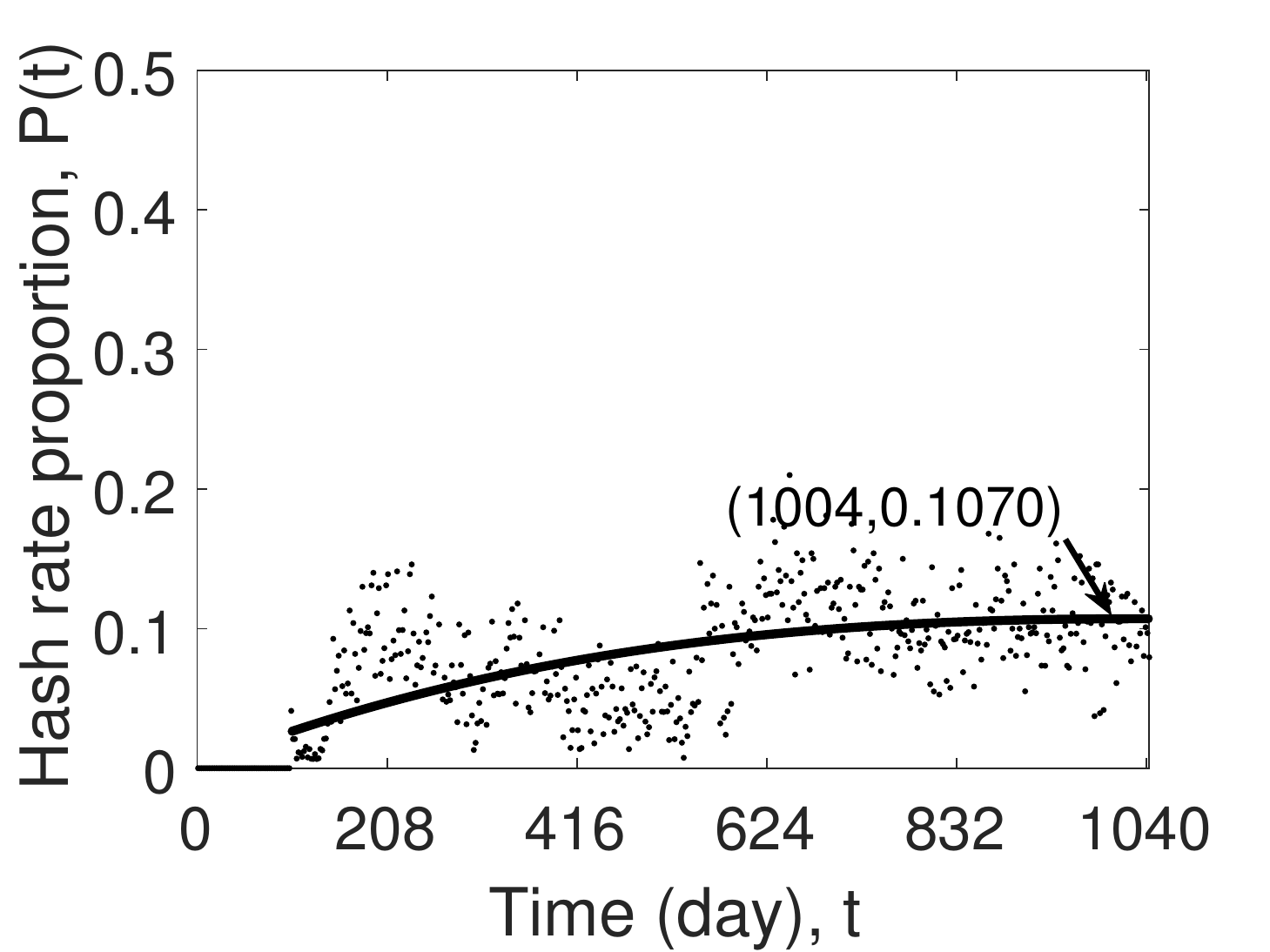}%
	}\hspace{0.2cm}
	\subfigure[BTC.com]{%
		\includegraphics[width=0.23\textwidth]{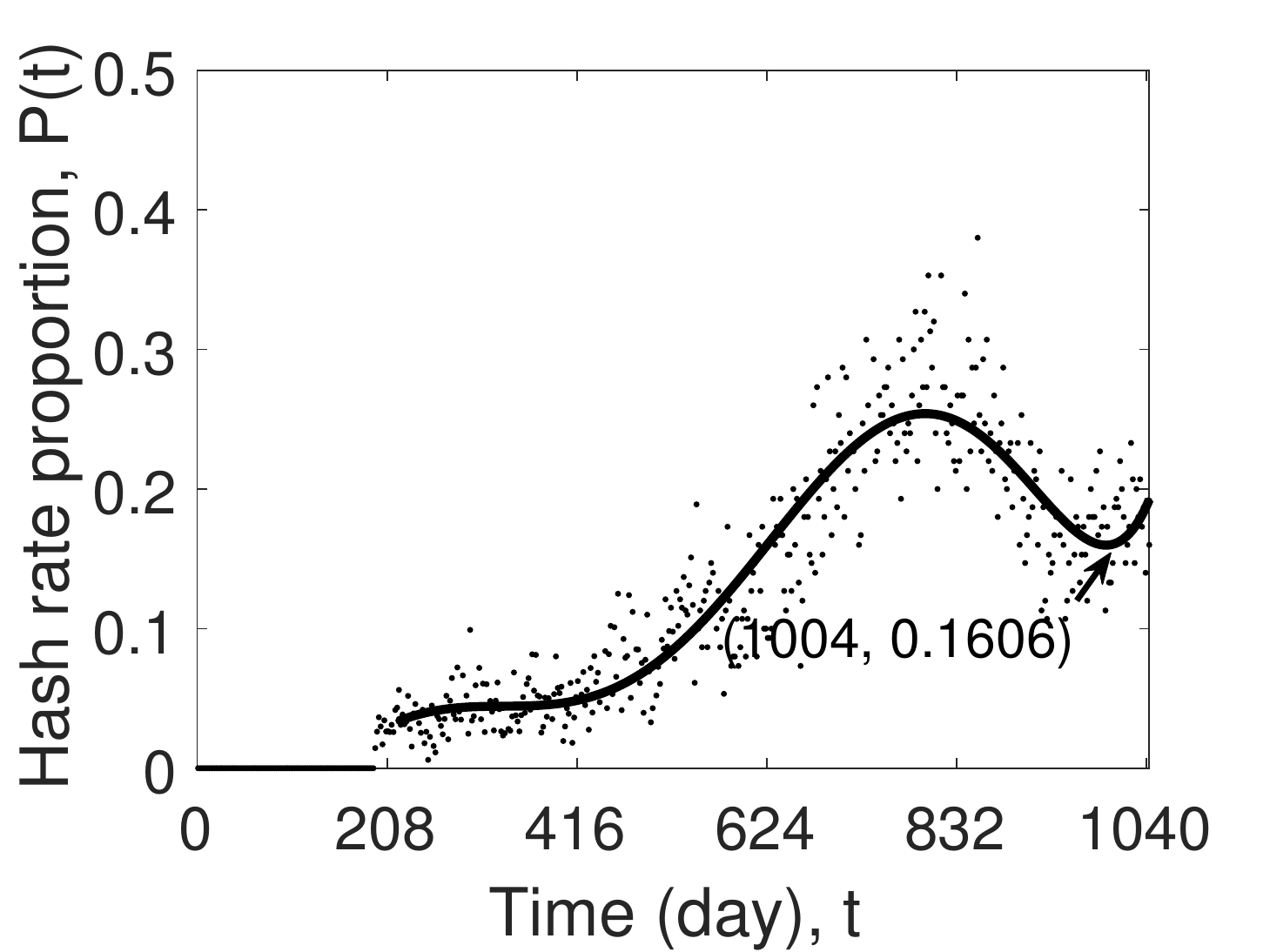}%
	}
	
	\caption{Daily Hash Rate Proportion of Top Mining Pools from Feb 25, 2016 to Jan 03, 2019}	
\end{figure*}

Driven by the Bitcoin incentive mechanism, mining pools contribute a large amount of computing power to the Bitcoin network. In the following section, we will dissect the computing power distribution and the evolution of Bitcoin mining pools. Data of computing power can be estimated based on the proportion of mining pool's blocks to the Bitcoin network's blocks over a period of time \cite{gencer2018decentralization}. Also, it can be obtained from the data servers such as Blockchain.Info and BTC.com \cite{pappalardo2018blockchain, wang2015exploring}. The advantage of the first method is that pruned blocks do not affect minted block distribution.

\subsubsection{Minted Block Distribution}

\begin{figure}[t]
	\centering
	
	\subfigure[]{%
		\includegraphics[width=0.23\textwidth]{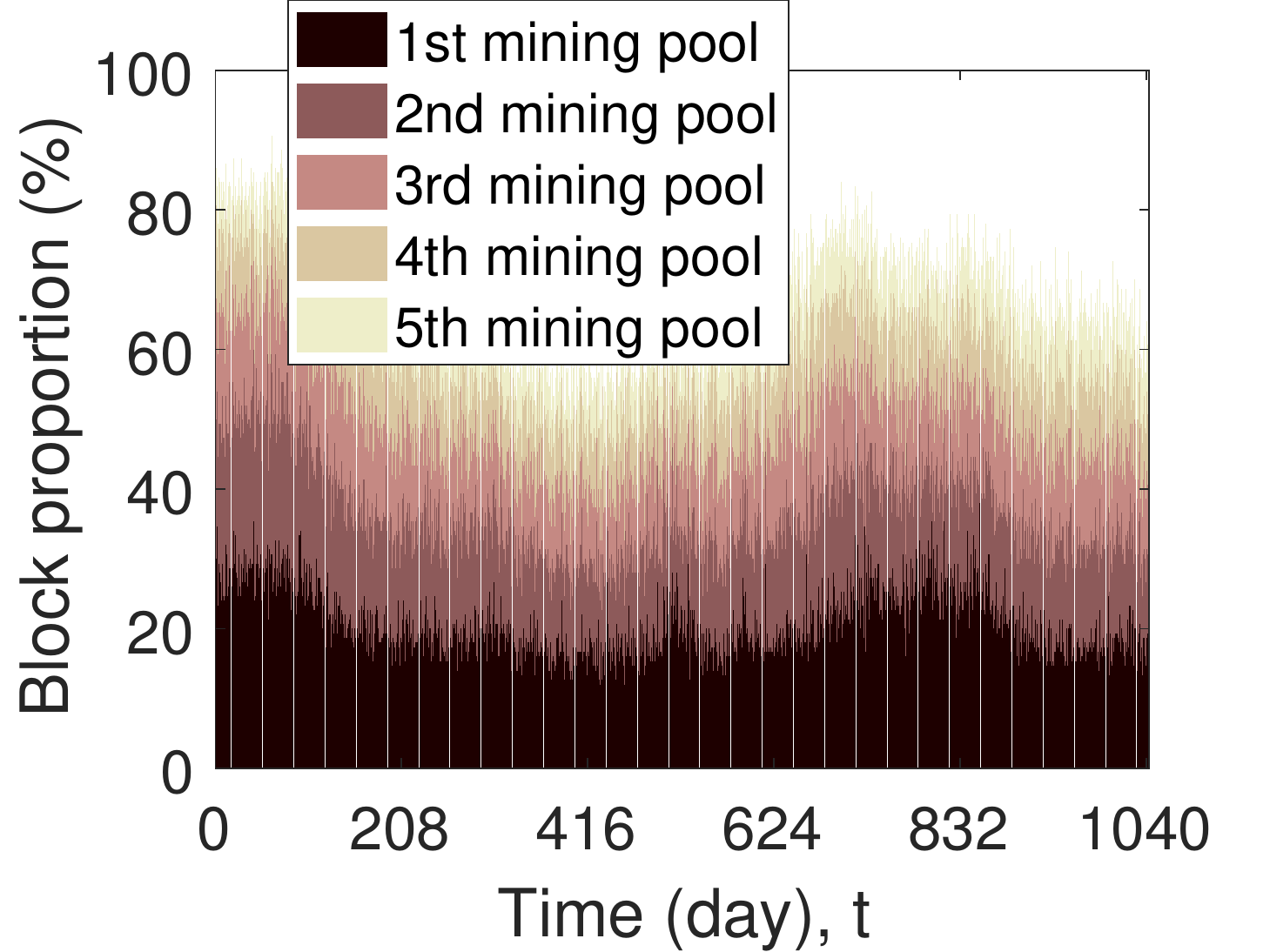}%
	}\hspace{0.2cm}
	\subfigure[]{%
		\includegraphics[width=0.23\textwidth]{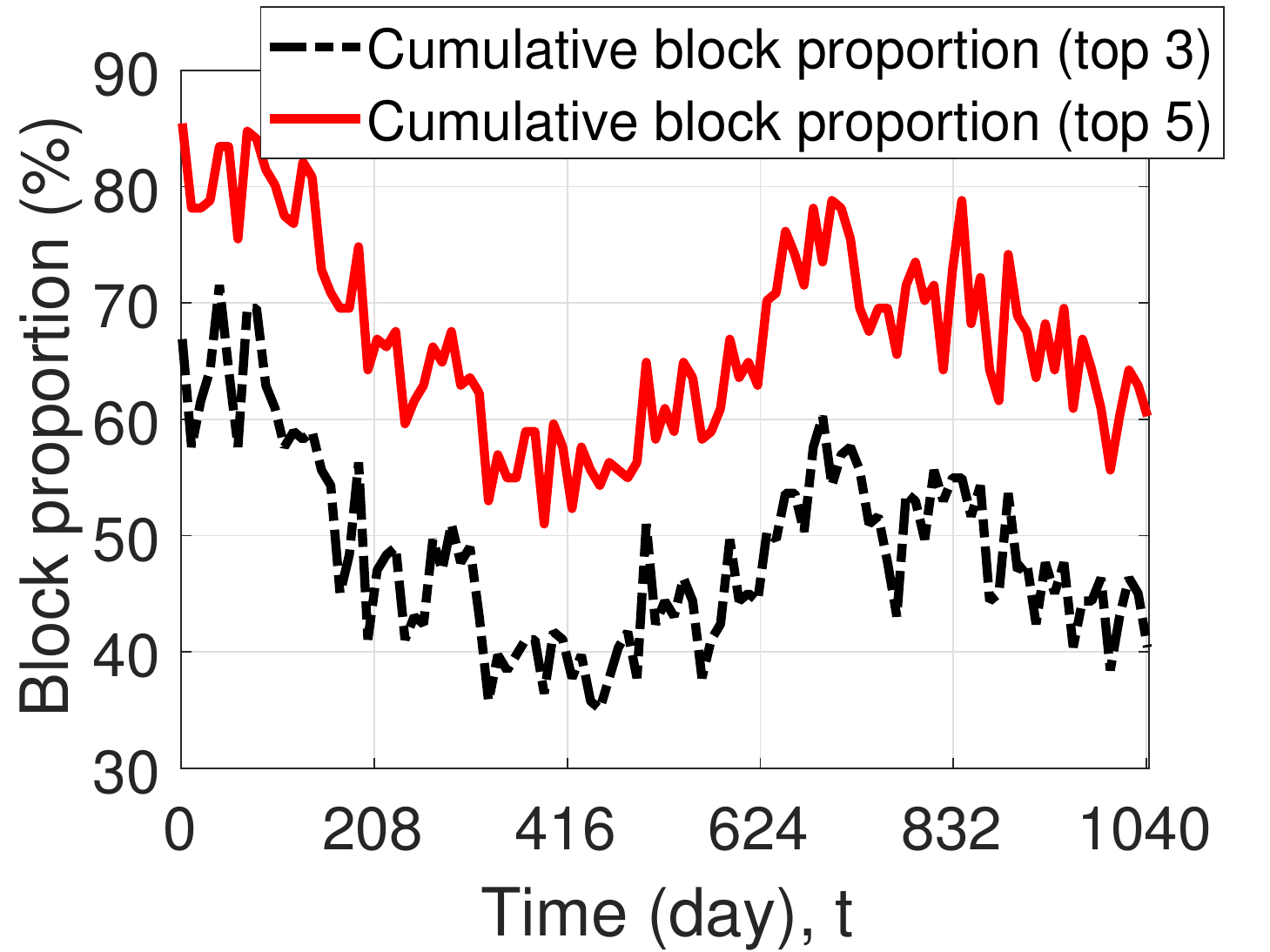}%
	}%
	
	\caption{(a) Daily Block Proportion of Top Mining Pools from Feb 25, 2016 to Jan 03, 2019; (b) Daily Cumulative Block Proportion of Top 3 and Top 5 Mining Pools from Feb 25, 2016 to Jan 03, 2019}
	
\end{figure}

Fig. 2 shows an estimation of blocks created by mining pools from Feb 25, 2016 to Jan 03, 2019, where the block proportion is calculated by dividing a specific mining pool's blocks by the Bitcoin network's blocks over a period of time. We found that over the 95.99 percent of the Bitcoin network's computing power is controlled by top 25 mining pools. Also, Fig. 2 shows an interesting \textit{phenomenon} that the Bitcoin network relies heavily on a few distinct mining pools \cite{gencer2018decentralization,gervais2014bitcoin}. During the observation period, the top four mining pools, i.e., AntPool, F2Pool, ViaBTC and BTC.com, created 48.44 percent of blocks; and top five mining pools, i.e., AntPool, F2Pool, ViaBTC, BTC.com and SlushPool, created 56.10 percent of blocks that has exceeded the 51\% computing power of the Bitcoin network.

To validate the \textit{phenomenon} of computing power centralization appears at any time during the observation period, we conducted experiments on block proportion for top mining pools. Fig. 5. (a) shows block proportion of top 1, 2, ..., 5 mining pool. Fig. 5. (b) shows cumulative block proportion of top 3 mining pools and top 5 mining pools. We found that more than 33 percent of the Bitcoin blocks are created by only top 3 mining pools entities every day. Similarly, more than 51 percent of the Bitcoin blocks are created by only top 5 mining pools entities every day. It means that top 3 mining pools entities continuously controlled more than 33 percent of daily computing power; and top 5 mining pools entities continuously controlled more than 51 percent of daily computing power.

\textit{The Security Concerns.} Although the Bitcoin protocol itself is purely decentralized, we found that the Bitcoin network relies heavily on major mining pools; and a few top mining pools entities continuously control most computing power every day. Such a trend of computing power centralization may raise security concerns such as the 51\% attacks and selfish mining attacks \cite{eyal2018majority} if these top mining pools aligned to create new blocks. Note that both 51\% attacks and selfish mining attacks are not just theoretical attacks. Many real world cases had been observed in other blockchain applications such as Namecoin\cite{ali2016blockstack, zheng2018blockchain}, Zencash and Bitcoin Gold.

\subsection{Hash Rate Distribution}

\begin{figure*}[t]
	\centering
	
	\subfigure[AntPool]{%
		\includegraphics[width=0.23\textwidth]{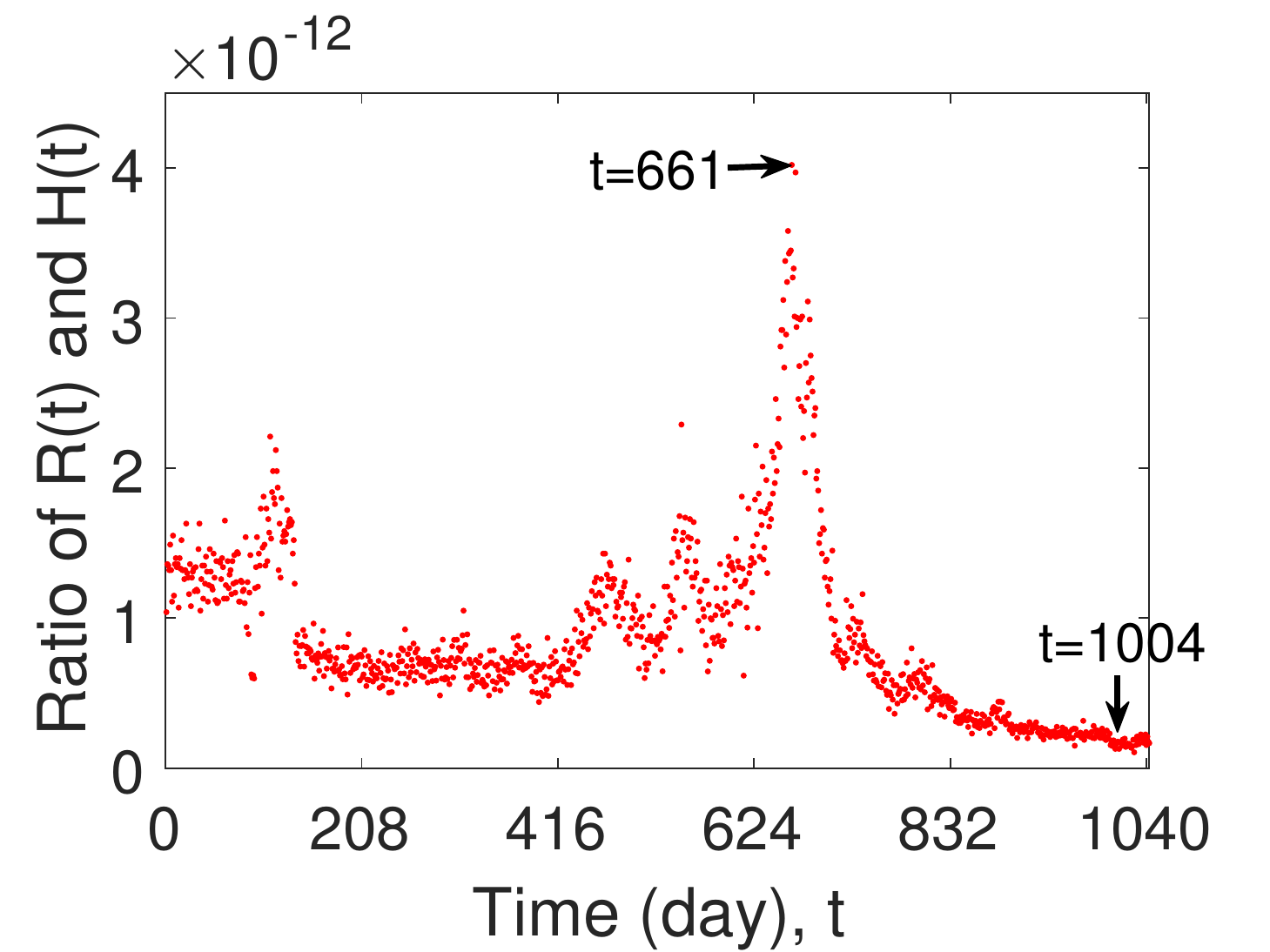}%
	}\hspace{0.2cm}
	\subfigure[F2Pool]{%
		\includegraphics[width=0.23\textwidth]{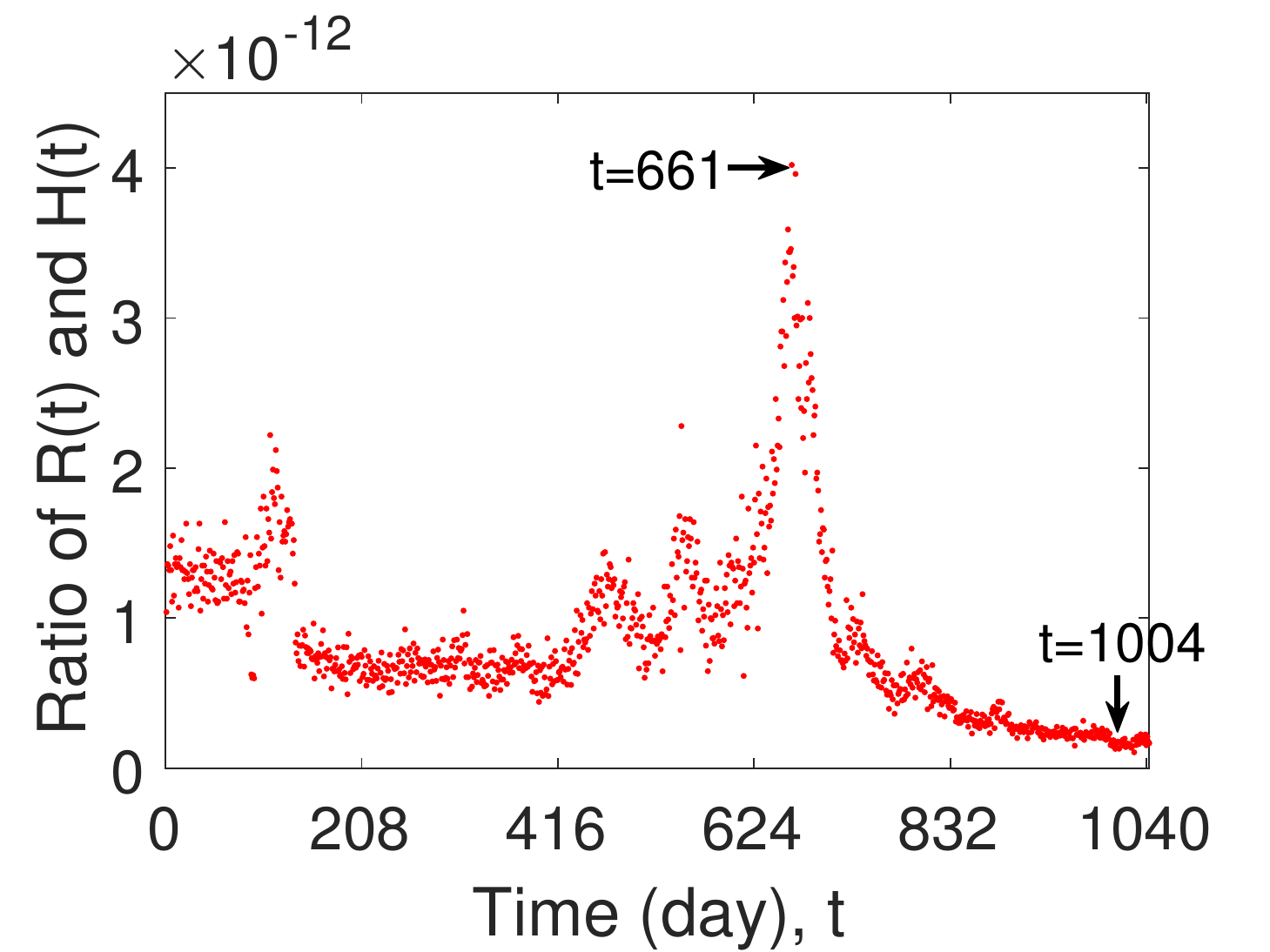}%
	}\hspace{0.2cm}
	\subfigure[ViaBTC]{%
		\includegraphics[width=0.23\textwidth]{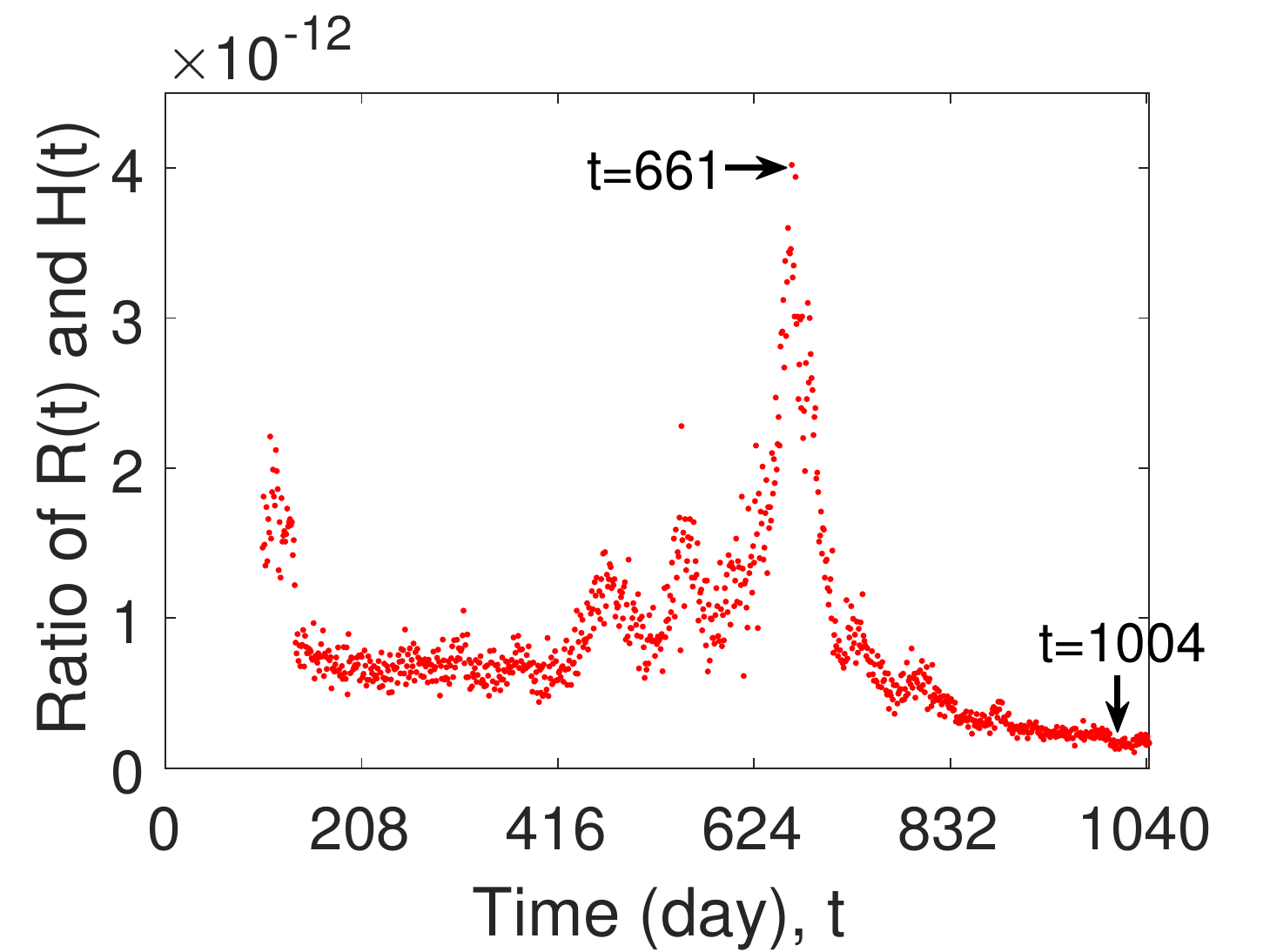}%
	}\hspace{0.2cm}
	\subfigure[BTC.com]{%
		\includegraphics[width=0.23\textwidth]{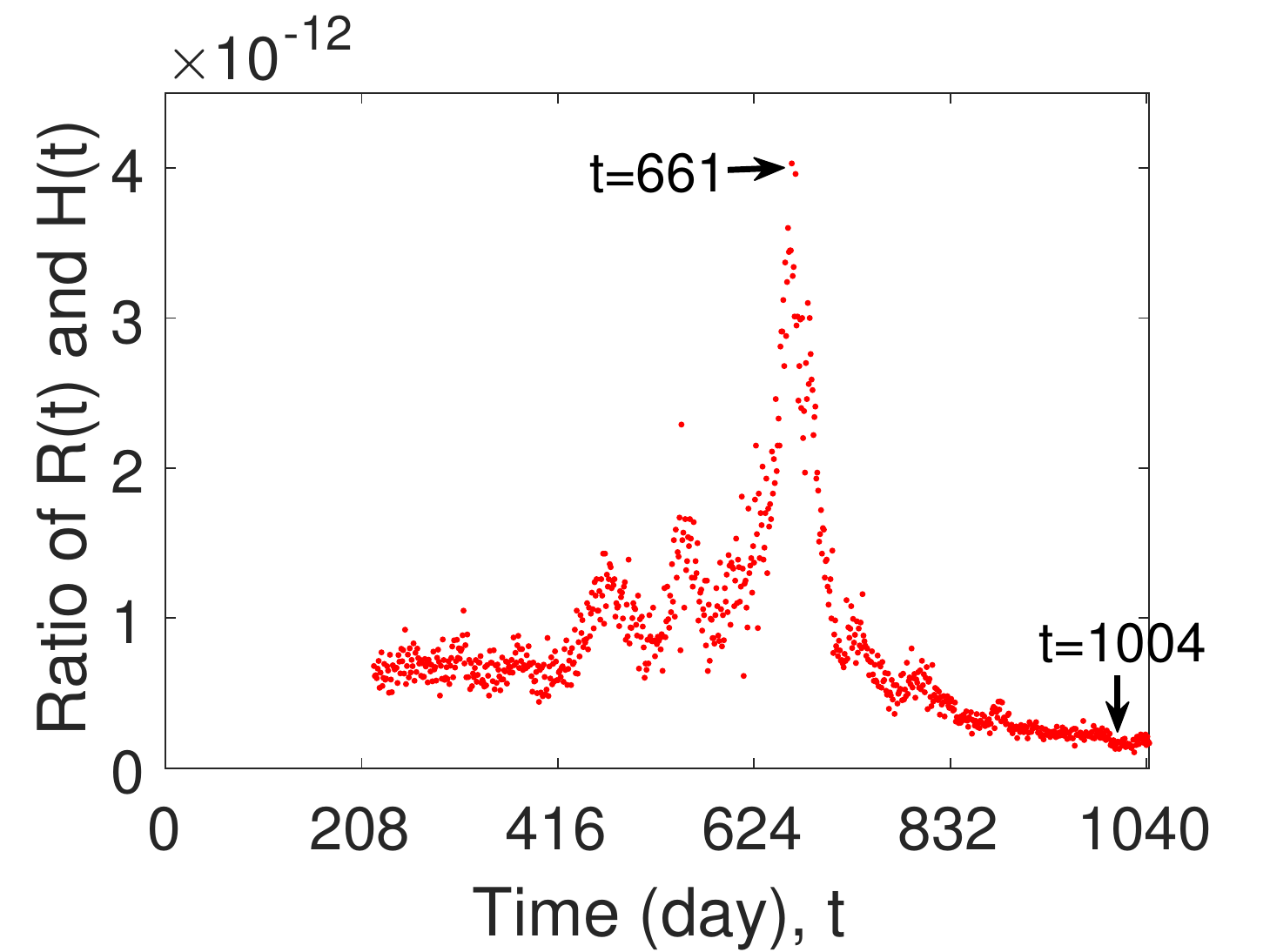}%
	}
	
	\caption{The Unit Profit of Mining Power (i.e., Hash Rate) from Feb 25, 2016 to Jan 03, 2019}	
\end{figure*}

The hash rate proportion is calculated by dividing the hash rate of a specific mining pool by the hash rate of the Bitcoin network over a period of time. Fig. 3 shows the daily hash rate of top mining pools from Feb 25, 2016, to Jan 03, 2019. For day $t \in \left [ 1,1004 \right ]$ when the hash rate of each mining pool continues to rise, we denote it by stage $S1$. For day $t \in \left [ 1004,1044 \right ]$ when the hash rate of each mining pool began to decline, we denote it by stage $S2$. 

We modeled the changes of the hash rate of mining pools in the first stage $S1$ by applying a simple Malthusian model. The Malthusian model is also called a simple exponential growth model, which is an exponential model with a constant growth rate. We denote time by $t$ and the hash rate of a mining pools at time $t$ by $H(t)$. We assume that the growth rate of the hash rate is a constant growing rate. That is, the hash rate $H(t)$ at any given time $t$; changes to $H(y+\delta t)$ at time $t+\delta t$, the growth rate $\lambda$ at time $t$ would be a constant value $\lambda = \frac{H(t+\delta t)-H(t)}{H(t)\delta t}$. Thus, based on the assumption, we have $\lambda = \lim_{\delta t\rightarrow 0}{\frac{H(t+\delta t)-H(t)}{H(t)\delta t}}$ by taking limit $\delta t\rightarrow 0$ as follows,

\begin{equation}
\lambda = \frac{H{}'(t)}{H(t)}
\end{equation}

After solving the math differential equation (1), we can model the hash rate $H(t)$ of a mining pool at any time $t$ as given in equation (2), where $\lambda$ is the constant growth rate and $c$ is a constant variable.

\begin{equation}
H(t)=e^{\lambda t}\cdot e^{c}
\end{equation}

To solve the variables in equation (2), we transform the equation into $ln(H(t))=\lambda x + c$, and then use the Least Square (LS) method to solve the variable $\lambda$ and $c$. TABLE 3 shows the parameters of $\lambda$ and $c$ of the hash rate model $H(t)=e^{\lambda t}\cdot e^{c}$. We found that as of stage $S1$, the Bitcoin network hash rate grows exponentially over time. Similarly, the hash rate of top mining pools grow exponentially during the observation period. Specifically, the hash rate of the Bitcoin network of stage $S1$ is $H(t)=e^{4.042 \times {10^-{3}}} \cdot e^{41.5198}$ and the hash rate of the AntPool of stage $S1$ is $H(t)=e^{3.561 \times {10^-{3}}} \cdot e^{39.9462}$.

\begin{table}[t]
	\centering
	\caption{Parameters of $\lambda$ and $c$ of the Hash Rate Model $H(t)=e^{\lambda t}\cdot e^{c}$}
	\scalebox{0.82}{
		\begin{tabular}{lllll}\hlinew{0.5pt}
			\hline
			Mining Entities & $\lambda$              & $c$     & R-square & Adj R-sq \\ \hline \hlinew{0.5pt}
			AntPool         & $3.561 \times 10^{-3}$ & 39.9462 & 0.8658   & 0.8657   \\
			F2Pool          & $4.429 \times 10^{-3}$ & 38.7316 & 0.8313   & 0.8311   \\
			ViaBTC          & $4.239 \times 10^{-3}$ & 39.1101 & 0.8403   & 0.8402   \\
			BTC.com         & $3.836 \times 10^{-3}$ & 40.0757 & 0.7966   & 0.7964   \\
			Bitcoin Network & $4.042 \times 10^{-3}$ & 41.5198 & 0.9483   & 0.9483   \\ \hline \hlinew{0.5pt}
		\end{tabular}
	}
\end{table}

\begin{figure*}[t]
	\centering
	
	\subfigure[Mining Rewards]{%
		\includegraphics[width=0.26\textwidth]{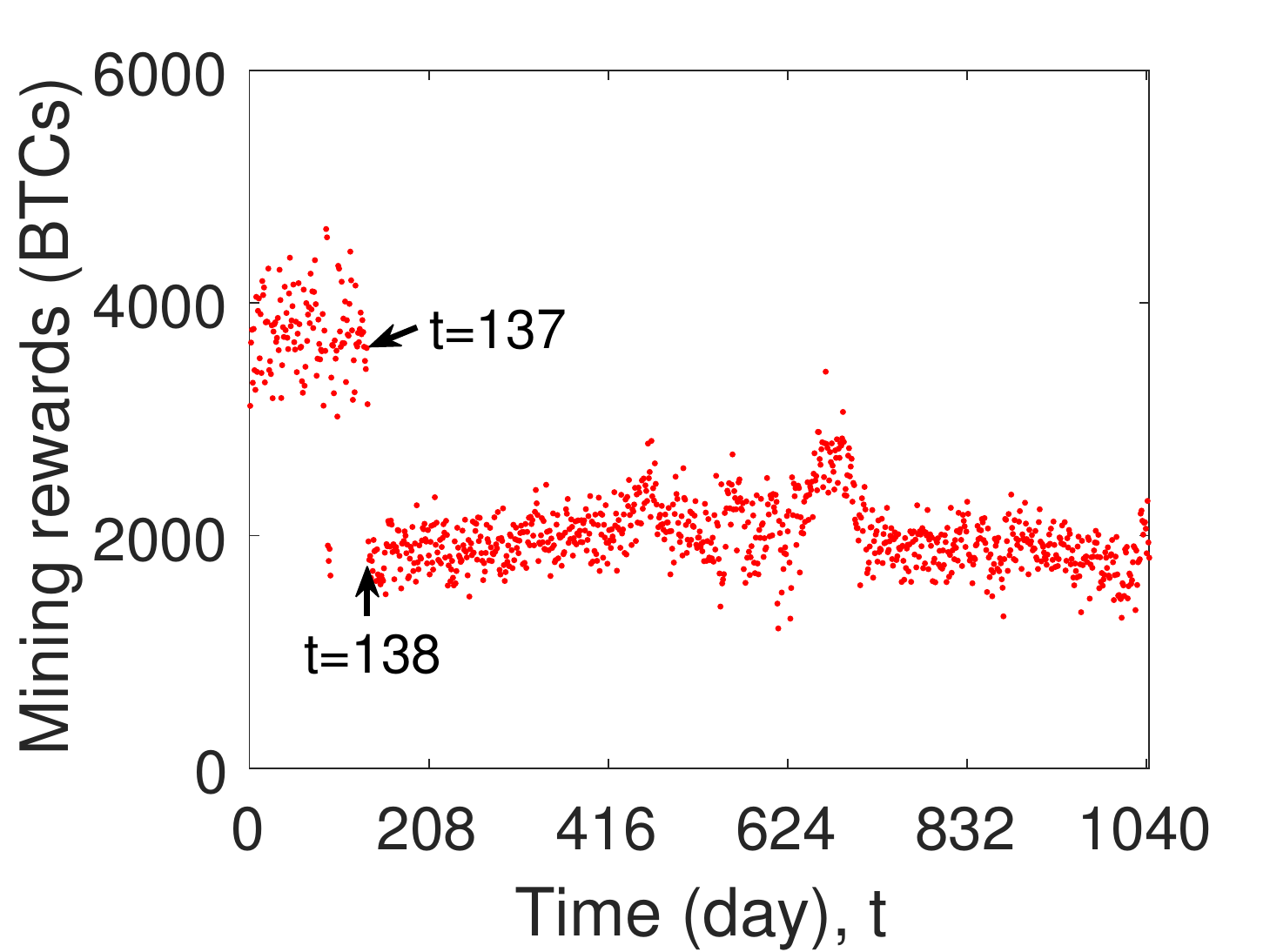}%
	}\hspace{0.2cm}
	\subfigure[Market Price]{%
		\includegraphics[width=0.26\textwidth]{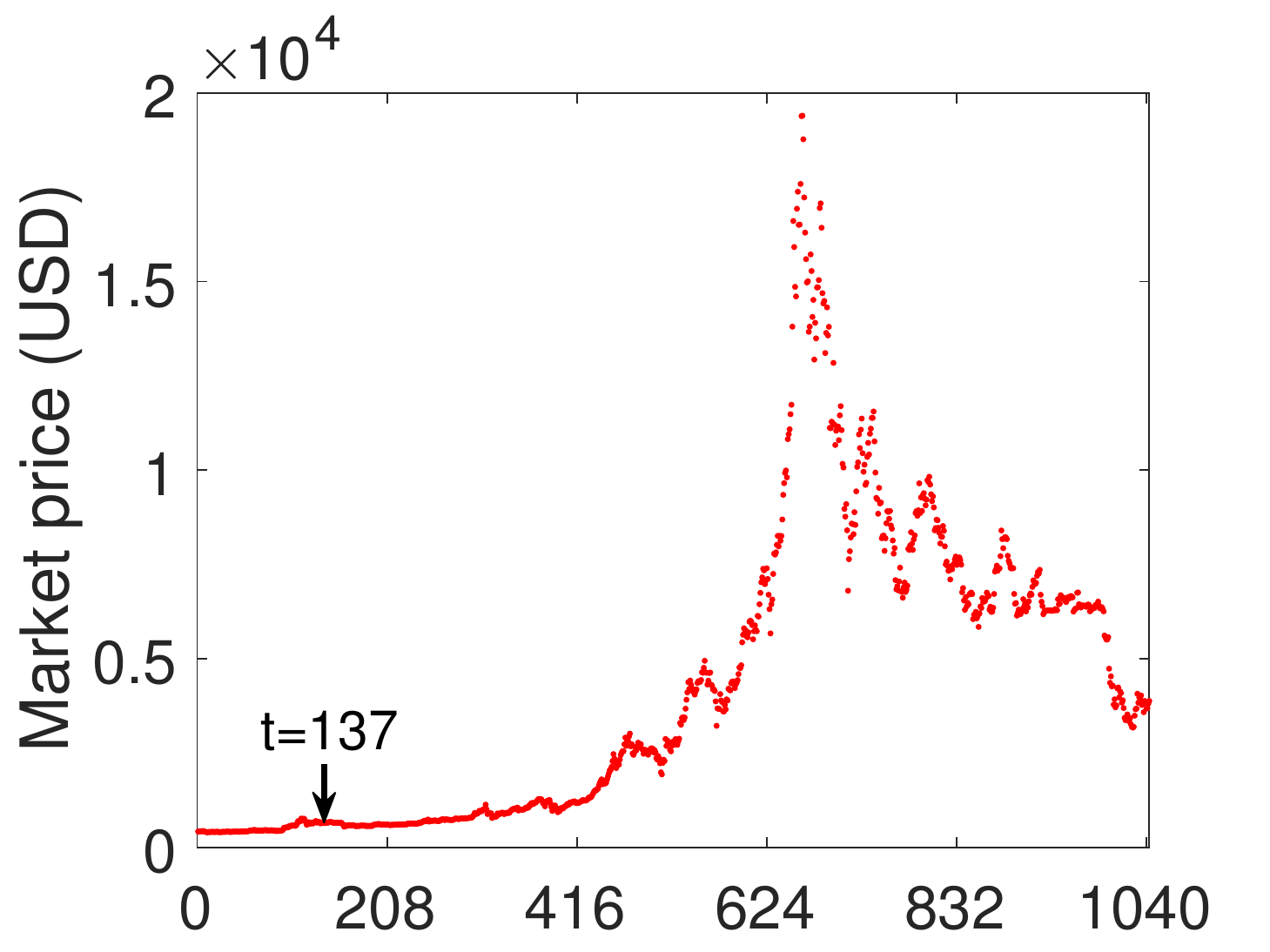}%
	}\hspace{0.2cm}
	\subfigure[Mining Revenue]{%
		\includegraphics[width=0.26\textwidth]{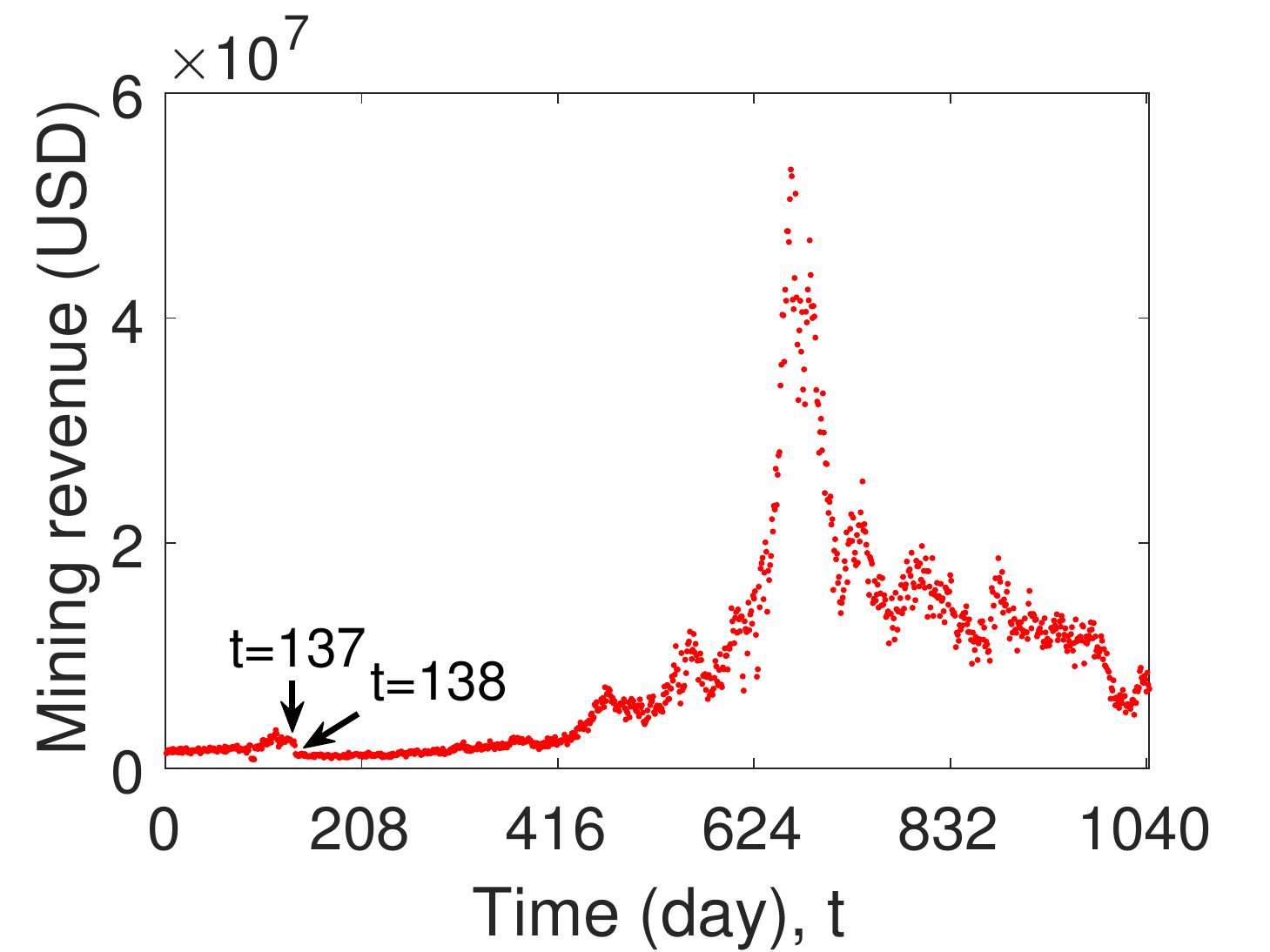}%
	}
	
	\caption{Daily Mining Rewards, Market Price and Mining Revenue from Feb 25, 2016 to Jan 03, 2019}	
\end{figure*}

\begin{table*}[t]
	\centering
	\caption{Game of Prisoner's Dilemma Among Two Mining Pools A and B}
	\begin{tabular}{ll|ll}
		&           & \multicolumn{2}{c}{{\normalsize Mining Pool A}} \\
		&           & {\normalsize Increase}        & {\normalsize Unchanged}        \\ \hline
		\multirow{2}{*}{{\normalsize Mining Pool B}} & {\normalsize Increase}  & $\left( \frac{ \alpha  \left(1+ \varepsilon  \right) }{\alpha  \left(1+\varepsilon \right)+ \beta  \left(1+\varepsilon \right)} , \frac{\beta \left( 1+ \varepsilon \right)}{\alpha  \left(1+\varepsilon \right)+ \beta  \left(1+\varepsilon \right)}  \right)  $               & $\left( \frac{ \alpha  }{\alpha + \beta  \left(1+\varepsilon \right)} , \frac{\beta \left( 1+ \varepsilon \right)}{\alpha  + \beta  \left(1+\varepsilon \right)}  \right)  $                \\
		& {\normalsize Unchanged} &$\left( \frac{ \alpha  \left(1+ \varepsilon  \right) }{\alpha  \left(1+\varepsilon \right)+ \beta  } , \frac{\beta }{\alpha  \left(1+\varepsilon \right)+ \beta  }  \right)$               &  $\left( \frac{ \alpha   }{\alpha  + \beta  } , \frac{\beta }{\alpha  + \beta  }  \right)$               
	\end{tabular}
\end{table*}

\textit{The Malthusian Trap}. We found an interesting \textit{phenomenon} that mining pools are caught in a Malthusian trap \cite{cilliers2018land, sugihara2019multiple}, where an exponential growth of the hash rate does not mean an increasing hash rate proportion of a mining pool. On the converse, some mining pools' hash rate proportions are actually decreasing despite an exponential growth of the hash rate. Fig. 4 shows the daily hash rate proportion of top mining pools from Feb 25, 2016 to Jan 03, 2019. Specifically, the hash rate proportion of AntPool dropped from 28.29 percent at day one to 11.58 percent at day 1004. Similarly, the hash rate proportion of F2Pool dropped from 26.29 percent at day one to 9.78 percent at day 1004, despite an exponential growth in the hash rate. We will further analyze the mining revenue and the prisoner's dilemma among mining pools in section 4.3.1, where we could find that mining pools are stuck in a Malthusian trap where there is a stage at which the Bitcoin incentives are inadequate for feeding the exponential growth of the computing resources.

\subsection{Mining Revenue}

The mining revenue is highly affected by Bitcoin market price and block reward transition. In this subsection, we will discuss the market price, block reward transition and their effects on mining pools. We developed a simple revenue model for mining pools. Assuming that the unit cost of computing power is a constant value denoted by $\upsilon$. And mining pools are rational, meaning that when mining activities are profitable mining pools will increase the hash rate. When mining activities are not profitable mining pools will decrease their hash rate. Let $t=1$ be Feb 25, 2016. We denote mining revenue of a given day $t$ by $R(t)$, mining rewards of a given day $t$ by $W(t)$ and the Bitcoin market price of a given day $t$ by $M(t)$. Note that the mining rewards $W(t)$ consist of two parts: transaction fees and block rewards. The transaction fees can be obtained by querying local blockchain data set. The block reward of a given $blockid$ can be calculated according to the Bitcoin protocol as follows,

\begin{equation}
50\times (1/2)^{\left \lfloor \frac{blockid}{210000} \right \rfloor}
\end{equation}

Assuming that both the mining rewards function $W(t)$ and the market price function $M(t)$ are continuous functions. Thus, given any time $t_1$ and $t_2$, the net mining profit, denoted by $Net$, can be calculated as follows,

\begin{equation}
Net(t_1,t_2) = \int_{t1}^{t2}{R(t)-\upsilon \cdot H(t)}dt
\end{equation}

\noindent where the mining rewards $R(t) = W(t) \cdot M(t)$. 

If we take $\Delta t = t_2 - t_1$ as a dairy datum. Then $R(t)$ is the average value of mining revenue per day and $H(t)$ is the average value of the hash rate of mining pools per day. Thus, we have the net profit of a mining pool of day $t$ given as follows,

\begin{equation}
Net(t)=R(t)-\upsilon \cdot H(t)
\end{equation}

\noindent where $\frac{R(t)}{H(t)}$ means the unit profit of hash rate. When $\upsilon \leq  \frac{R(t)}{H(t)}$, it means that mining activities for mining pools are profitable and mining pools will increase their hash rate. And when $\upsilon > \frac{R(t)}{H(t)}$, it means that mining activities for mining pools are not profitable and mining pools will not increase their hash rate.

We used our model to explain mining revenues of mining pools and the hash rate evolution. Fig. 6 shows the unit profit of mining power (i.e., $R(t)/H(t)$) from Feb 25, 2016 to Jan 03, 2019. We found that when $t=1004$, the unit profit of mining power dropped to a low value around $1.5 \times 10^{-13}$, of which hash rate of top mining pols began to decline in order to avoid further decline in unit profit of mining power. But when the hash rate of mining pools drops to a certain low value, the competitiveness of mining is weakened, and the unit profit of mining power will increase again. For example, on day $1034$ the unit profit of mining power rebounded to $2.3 \times 10^{-13}$.

An interesting \textit{phenomenon} is shown by combining Fig. 3 and Fig. 6 When day $t \in \left [ 661, 1004 \right ]$, unit profit of mining power is declining but the hash rates of mining pools are still growing exponentially. For example, for AntPool, the unit profit of mining power dropped from $4.02 \times 10^{-12}$ of day 661 to $0.15\times 10^{-12}$ of day 1004, but the hash rate of AntPool is still growing exponentially. We are interested in this \textit{\textbf{phenomenon}} and propose a simple model based on the game theory as shown in TABLE 4.

\subsubsection{Game of Prisoner's Dilemma among Mining Pools}

In order to analyze the \textit{phenomenon} as shown in Fig. 6 from day 661 to day 1004, where unit profit of mining power is declining but the hash rates of mining pools are still growing exponentially. We propose a simple model as follows. Suppose two mining pools A and B with initial computing power of $\alpha$ and $\beta$, respectively. Assume each mining pool increases its computing power at a growth rate $\varepsilon$ where $\varepsilon  \geq 0$. Each mining pool is rationally driven by incentive mechanism and maintains their existing market share by investing computing power to keep up with Bitcoin mining market. In our setting, two possible actions are considered: increase in computing power or keep computing power unchanged. According to the pay off matrix as shown in TABLE 4, the outcome of this game will be $\left( \frac{ \alpha  \left(1+ \varepsilon  \right) }{\alpha  \left(1+\varepsilon \right)+ \beta  \left(1+\varepsilon \right)} , \frac{\beta \left( 1+ \varepsilon \right)}{\alpha  \left(1+\varepsilon \right)+ \beta  \left(1+\varepsilon \right)}  \right) $ because this is the optimal strategy for each individual mining pool and no one has the motivation to change their strategy, although the ideal outcome for both A and B is $\left( \frac{ \alpha   }{\alpha  + \beta  } , \frac{\beta }{\alpha  + \beta  }  \right)$. To better understand the game of prisoner's dilemma among mining pools, we consider the following scenarios:

\begin{itemize}
	
	\item If mining pool A chooses to increase its computing power at a rate $\varepsilon$, then mining pool B will also increase its computing power at a rate of $\varepsilon$ or even higher, since B's mining revenue will drop from $\frac{\beta }{\alpha  + \beta  }$ to $\frac{\beta }{\alpha  \left(1+\varepsilon \right)+ \beta  }$ if it does not increase its computing power. 
	
	\item If mining pool A chooses to keep computing power unchanged, then mining pool B will also choose to increase its computing power in this case at a growth rate of $\varepsilon$, since B's mining revenue will increase from $\frac{\beta }{\alpha  + \beta  }$ to $\frac{\beta \left( 1+ \varepsilon \right)}{\alpha  + \beta  \left(1+\varepsilon \right)}$.
	
\end{itemize}

Thus, regardless of whether mining pool A increases or keeps its computing power unchanged, mining pool B will always increase its computing power. Due to the symmetry of the pay off matrix, mining pool A will also increase its computing power regardless whether mining pool B increases its computing power or not. Therefore, if mining pools are not cooperatively mining in a pool (see section 4.1.1), the final outcome of the game will be $\left( \frac{ \alpha  \left(1+ \varepsilon  \right) }{\alpha  \left(1+\varepsilon \right)+ \beta  \left(1+\varepsilon \right)} , \frac{\beta \left( 1+ \varepsilon \right)}{\alpha  \left(1+\varepsilon \right)+ \beta  \left(1+\varepsilon \right)}  \right) $, which results in a fact that even if mining revenues of mining pools decrease, game of computing power among mining pools are still fierce.

\subsubsection{Block Rewards Transition}

Bitcoin provides two incentives for miners: block rewards and transaction fees. The block reward dominates the incentive mechanism in the early days of the Bitcoin network. With the halving of block rewards, the transaction fees will gradually increase its proportion in the incentive mechanism. The assumption of Bitcoin is that no matter miners are paid by block rewards or transaction fees, it will not affect the Bitcoin incentive mechanism, and will not further affect the security of the Bitcoin network \cite{carlsten2016instability}. In this part, we are interested in the block reward transition and how it effects on mining revenue.

Fig. 7 shows the daily data of mining rewards, Bitcoin market price and mining revenue from Feb 25, 2016 to Jan 03, 2019. We found an interesting \textit{phenomenon} that the transaction fees and Bitcoin market prices are not sensitive to the event of halving block rewards. To be specific, TABLE 5 shows all two events of halving block rewards in the Bitcoin network. For the most recent event of halving block rewards, we can see that transaction fees slightly decreased from 46.6 BTCs on July 9, 2016 (GMT), to 38.3 BTCs on July 10, 2016 (GMT). Similarly, market price slightly decreased from \$649.96 on July 9, 2016 (GMT), to \$649.03 on July 10, 2016 (GMT). For the first event of halving block rewards on November 28, 2012 (GMT), it also shows the similar \textit{phenomenon} in the block reward transition.

\textit{Concerns on the Block Reward Transition.} The incentive mechanism including block rewards and transaction fees, contributes to the network security. According to the original design \cite{nakamoto2008bitcoin}, Bitcoin block rewards will eventually be cut half to zero. Until that time, the incentive mechanism will entirely rely on transaction fees. An implicit belief of the original design is that whether miners are rewarded with block rewards or transaction fees, will not affect the security of the Bitcoin network \cite{carlsten2016instability}.

Our measurement results show that both transaction fees and Bitcoin market prices are not sensitive to the block reward transition. As shown in Fig. 7 and TABLE 5, both transaction fees and market prices, cannot compensate for the half of block rewards. We suggested that a relatively stable incentive mechanism is of great importance to the security of the Bitcoin network, especially in a real world market where Bitcoin price is fluctuating.

\begin{table}[t]
	\centering
	\caption{All Two Events of Halving Block Rewards in the Bitcoin Network (until Jan 03, 2019)}
	\scalebox{0.68}{
		\begin{tabular}{lllll}\hlinew{0.5pt}
			\hline
			\multirow{2}{*}{\begin{tabular}[c]{@{}l@{}}Time\\ (mm/dd/yy)\end{tabular}} & \multicolumn{2}{l}{Mining Rewards (BTCs)} & \multirow{2}{*}{\begin{tabular}[c]{@{}l@{}}Market Price\\ (USD)\end{tabular}} & \multirow{2}{*}{\begin{tabular}[c]{@{}l@{}}Mining Revenue\\ (USD)\end{tabular}} \\  
			& Block rewards      & Transaction fees     &                                                                               &                                                                                 \\ \hline \hlinew{0.5pt}
			11/28/2012                                                                 & \textit{7,972}                & 33.8                 & 12.35                                                                         & \textit{98,872}                                                                          \\
			11/29/2012                                                                 & \textit{4,174}               & 30.7                 & 12.45                                                                         & \textit{52,349}                                                                          \\
			07/09/2016                                                                 & \textit{3,086}               & 46.6                 & 649.96                                                                        & \textit{2,036,064}                                                                       \\
			07/10/2016                                                                       & \textit{1,914}               & 38.3                 & 649.03                                                                        & \textit{1,267,101}                                                                       \\ \hline \hlinew{0.5pt}
		\end{tabular}
	}
\end{table}

\subsection{Block Size}

Block size is important to the Bitcoin performance \cite{matetic2018bite}. The maximum throughput of the Bitcoin network is constrained by block size and block interval \cite{croman2016scaling}. For example, a block size of 1 MB can support a maximum throughput of 3.3 to 7 transactions per second. Given the importance of block size, in this part we will explore the block size of different mining pools, with particular interest in the empty block analysis.

Fig. 8 shows the block size of mining pools from Feb 25, 2016 to Jan 3, 2019. We found that the block size of top mining pools is around 1 MegaByte (MB). For example, the average block size of AntPool is around 0.81 MB. And the average block size of BTC.com is around 0.90 MB.

Fig. 8 also shows an interesting \textit{phenomenon} that mining pools prefer 1 MB blocks to 2 MB blocks even though mining pools support a maximum block size of 2 MB following the Bitcoin improvement proposal 102 on November 11, 2015. For example, 95 percent of AntPool's blocks are less than 1.14 MB; 95 percent of F2Pool's blocks are less than 1.15 MB; 95 percent of ViaBTC's blocks are less than 0.95 MB and 95 percent of BTC.com's blocks are less than 1.23 MB. We are interested in why mining pools with the capability of generating the 2 MB block, perfer to generate 1 MB blocks. More analysis about transaction and block propagation will be discussed in section 5.1. 

\begin{figure}[t]
	\centering
	
	\subfigure[AntPool]{%
		\includegraphics[width=0.23\textwidth]{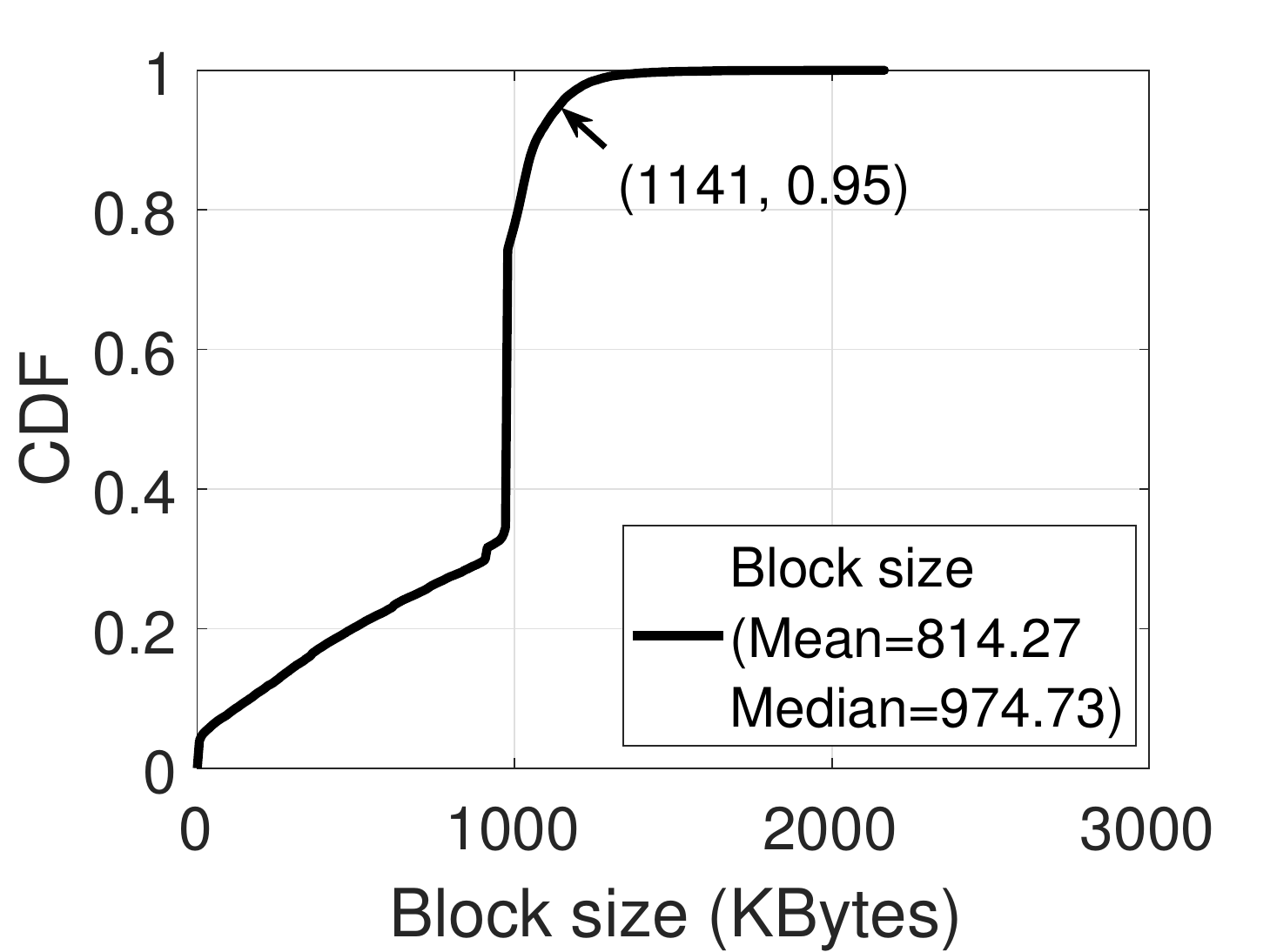}%
	}\hspace{0.2cm}
	\subfigure[F2Pool]{%
		\includegraphics[width=0.23\textwidth]{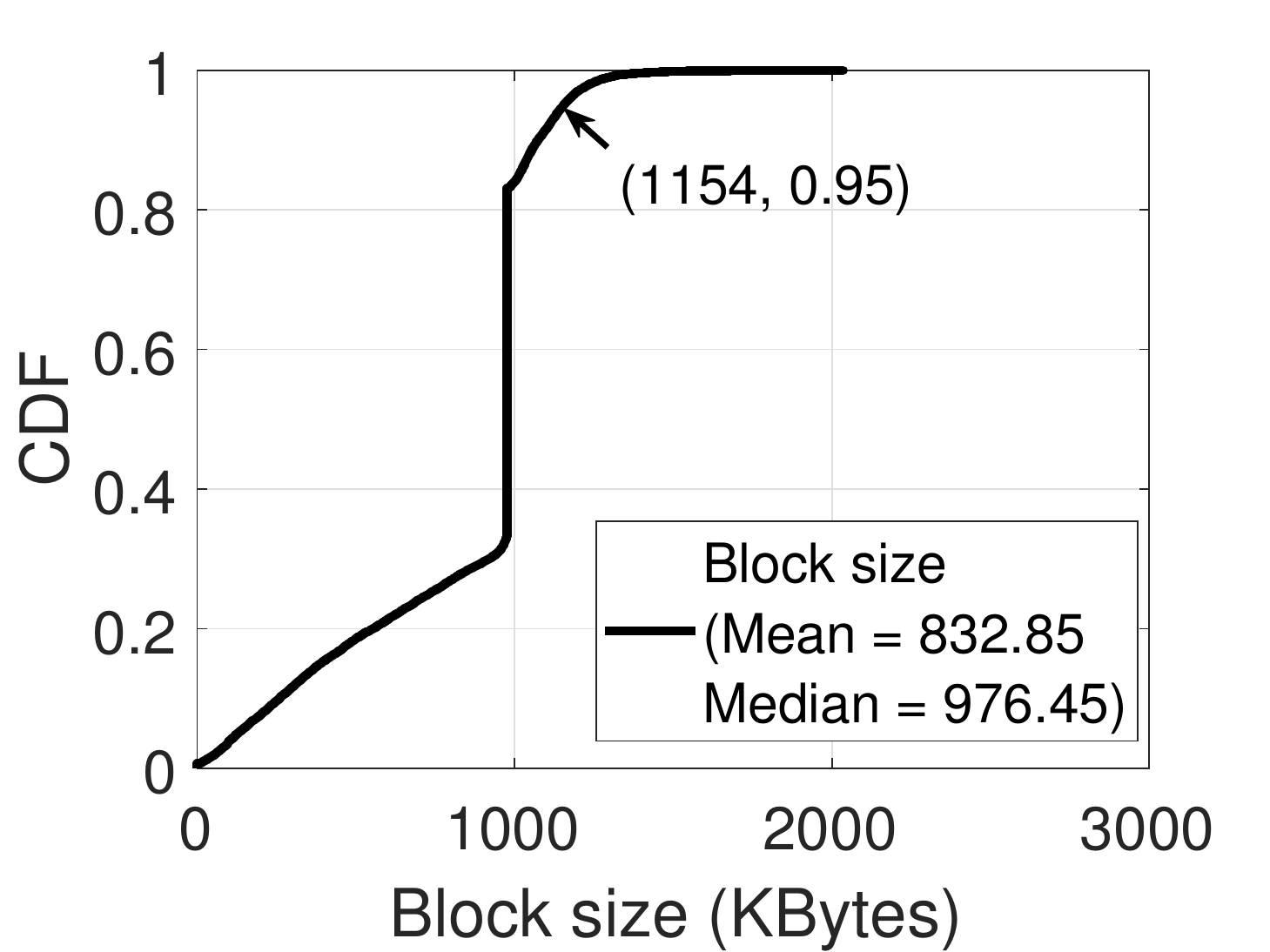}%
	}%
	
	\subfigure[ViaBTC]{%
		\includegraphics[width=0.23\textwidth]{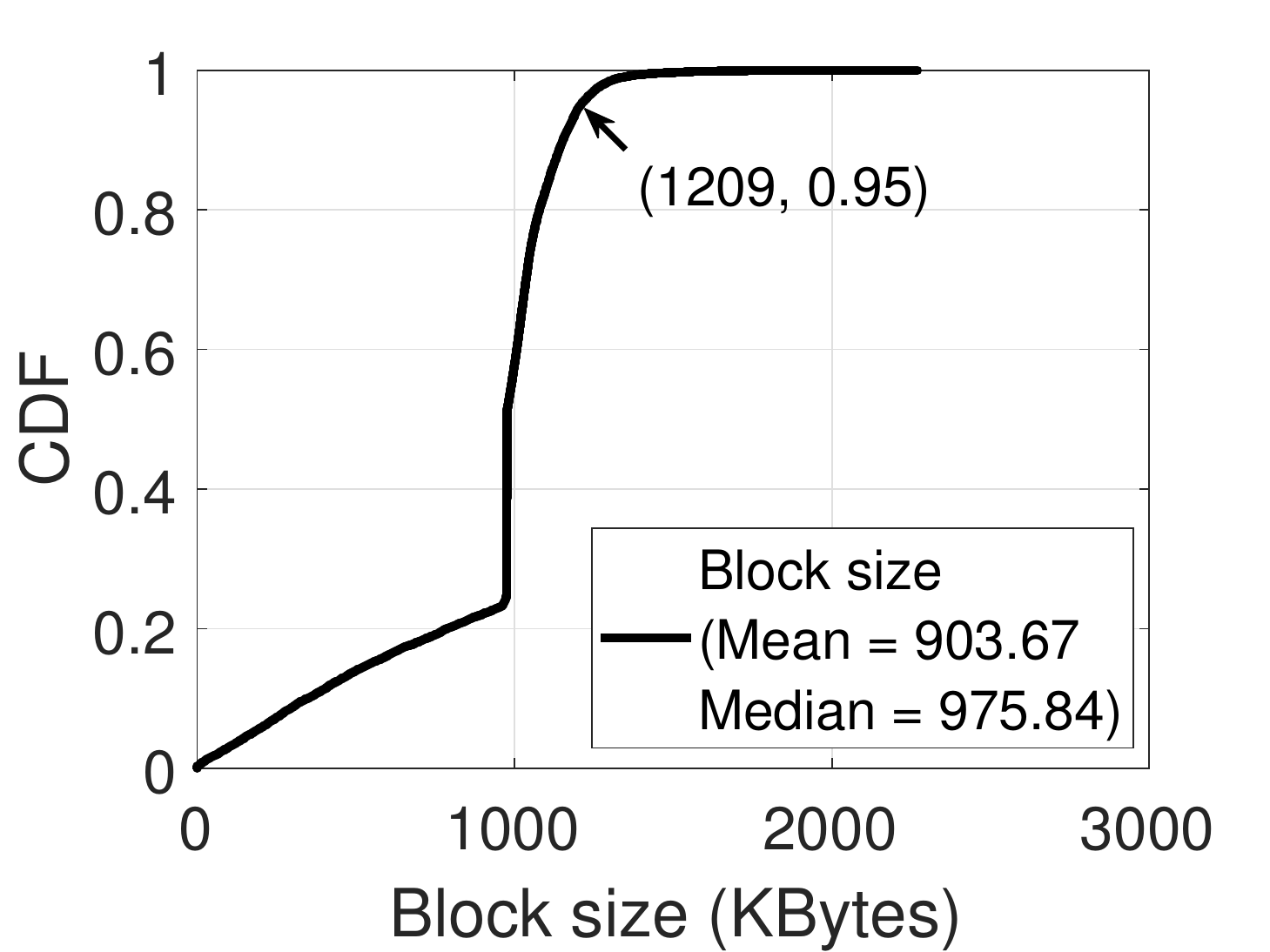}%
	}\hspace{0.2cm}
	\subfigure[BTC.com]{%
		\includegraphics[width=0.23\textwidth]{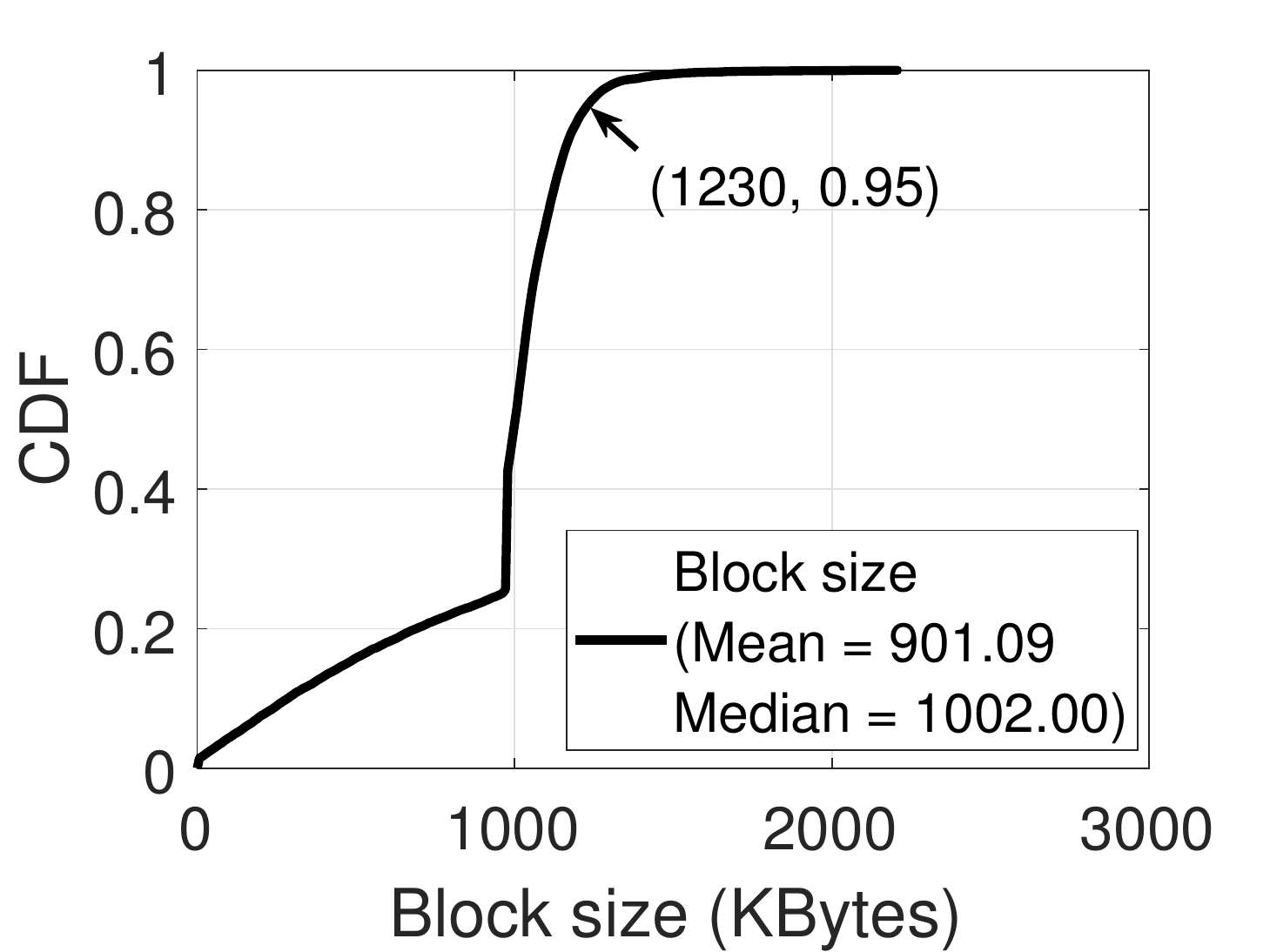}%
	}%
	
	\caption{Block Size of Mining Pools from February 25, 2016 to January 3, 2019}
\end{figure}

\begin{table}[t]
	\centering
	\caption{Empty Block Rates of Mining Pools from February 25, 2016 to January 3, 2018}
	\scalebox{0.73}{
		\begin{tabular}{llllll}\hlinew{0.5pt}
			\hline
			\begin{tabular}[c]{@{}l@{}}Mining Pools \\ Entities\end{tabular} & \begin{tabular}[c]{@{}l@{}}Bitcoin\\ network\end{tabular} & AntPool & F2Pool & ViaBTC & BTC.com \\ \hline \hlinew{0.5pt}
			The empty block rate                                                 & 1.01\%                                                    & 1.98\%  & 0.65\% & 0.26\% & 1.31\%  \\ \hline \hlinew{0.5pt}
		\end{tabular}
	}
\end{table}

\begin{figure*}[t]
	\centering
	
	\subfigure[Creat New Transactions]{%
		\includegraphics[width=0.22\textwidth]{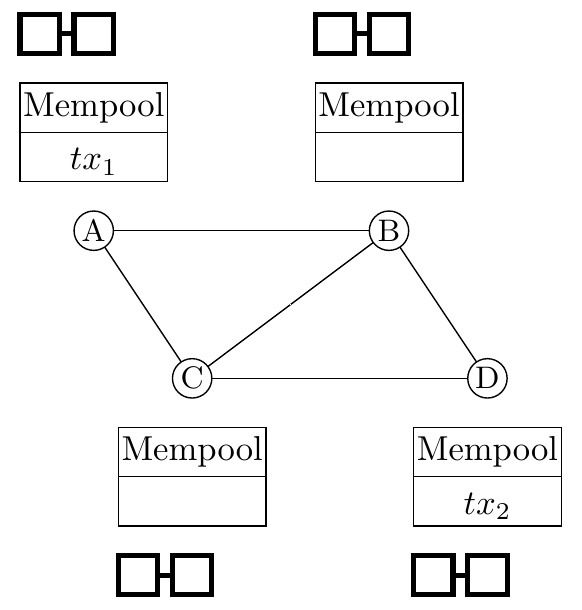}%
	}\hspace{0.20cm}
	\subfigure[Broadcast New Transactions]{%
		\includegraphics[width=0.22\textwidth]{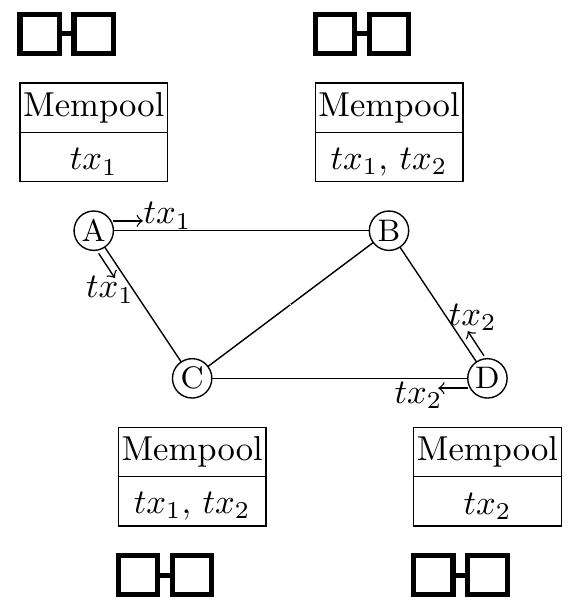}%
	}\hspace{0.20cm}
	\subfigure[Create a New Block]{%
		\includegraphics[width=0.22\textwidth]{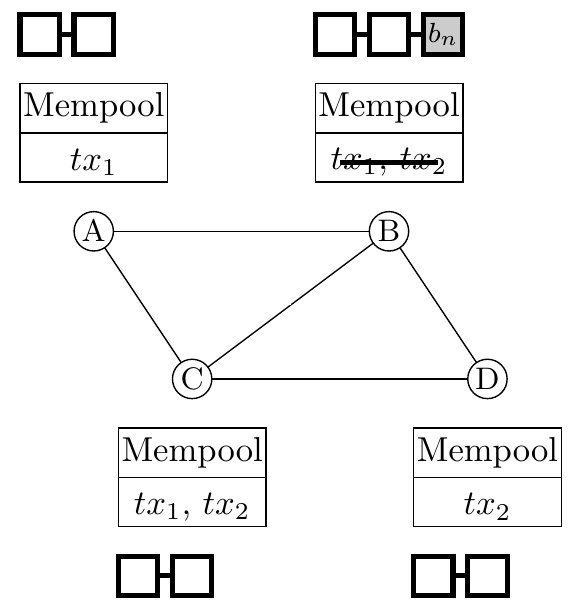}%
	}\hspace{0.20cm}
	\subfigure[Broadcast the New Block]{%
		\includegraphics[width=0.22\textwidth]{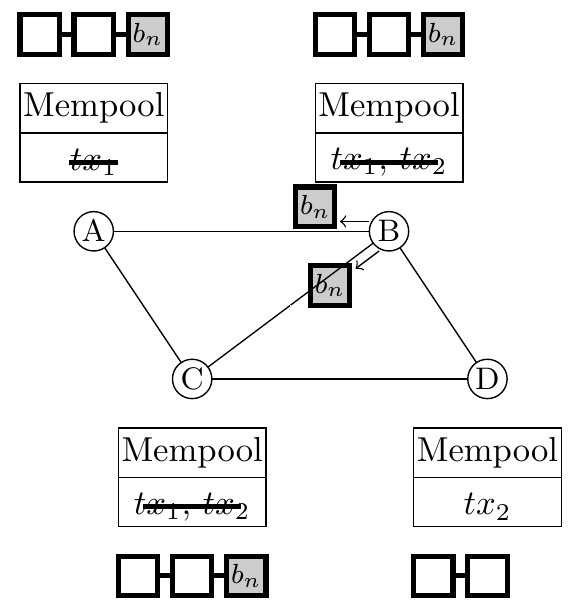}%
	}
	
	\caption{The Process of Broadcasting Transactions and Blocks in the Bitcoin Network (via the Bitcoin protocol)}	
\end{figure*}

\begin{figure}[t]
	\centering
	
	\subfigure[AntPool]{%
		\includegraphics[width=0.23\textwidth]{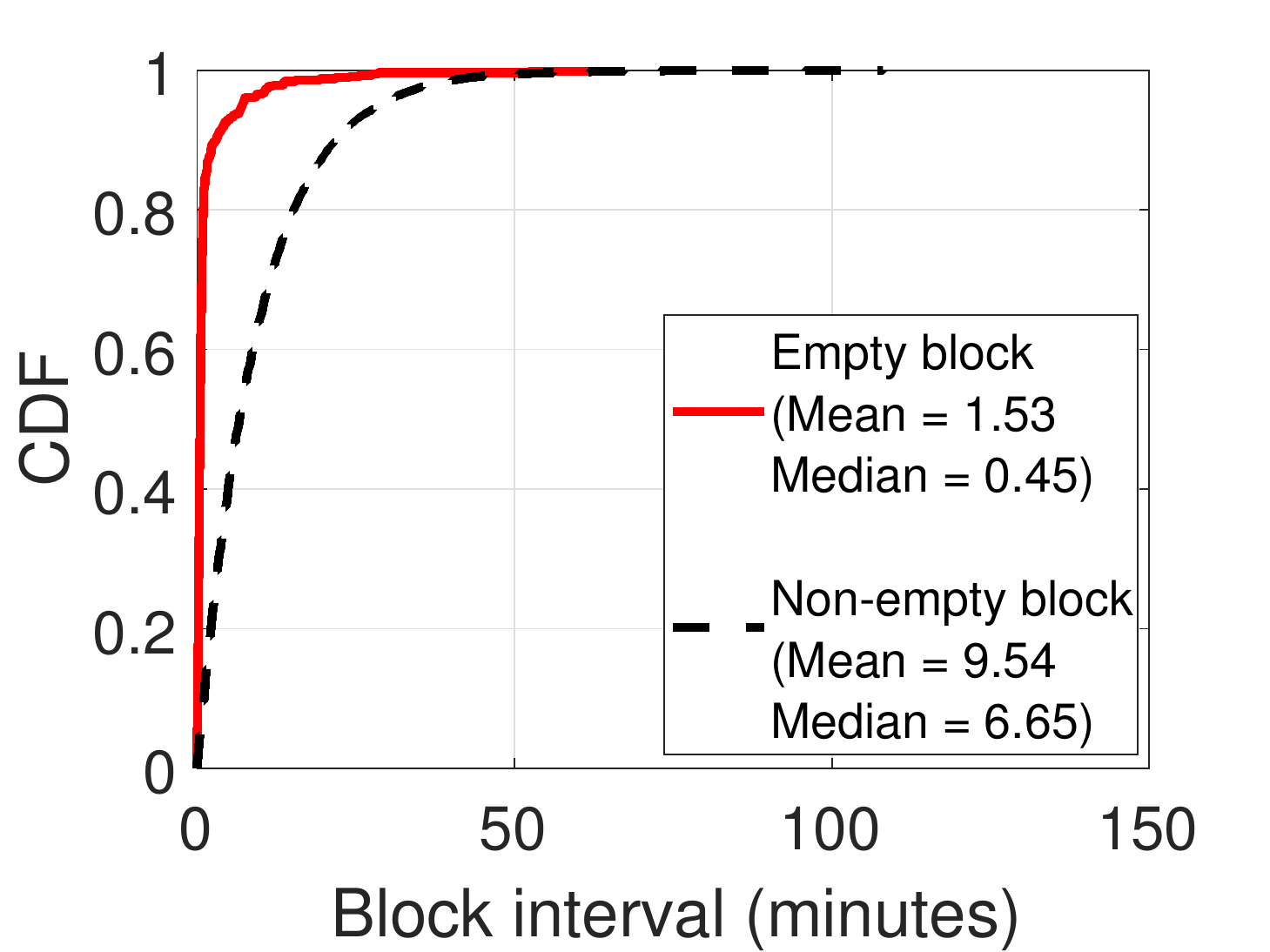}%
	}\hspace{0.2cm}
	\subfigure[F2Pool]{%
		\includegraphics[width=0.23\textwidth]{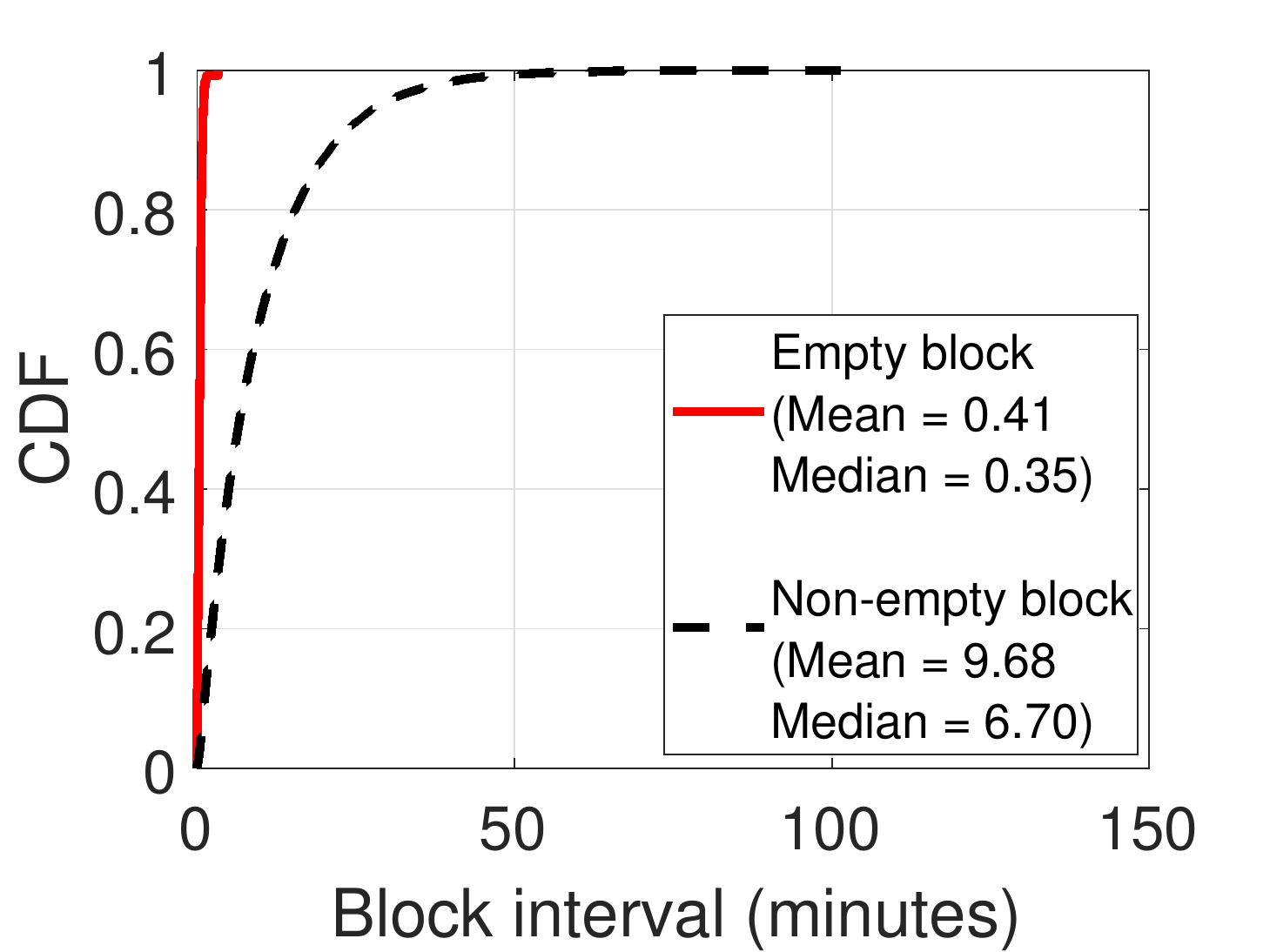}%
	}%
	
	\subfigure[SlushPool]{%
		\includegraphics[width=0.23\textwidth]{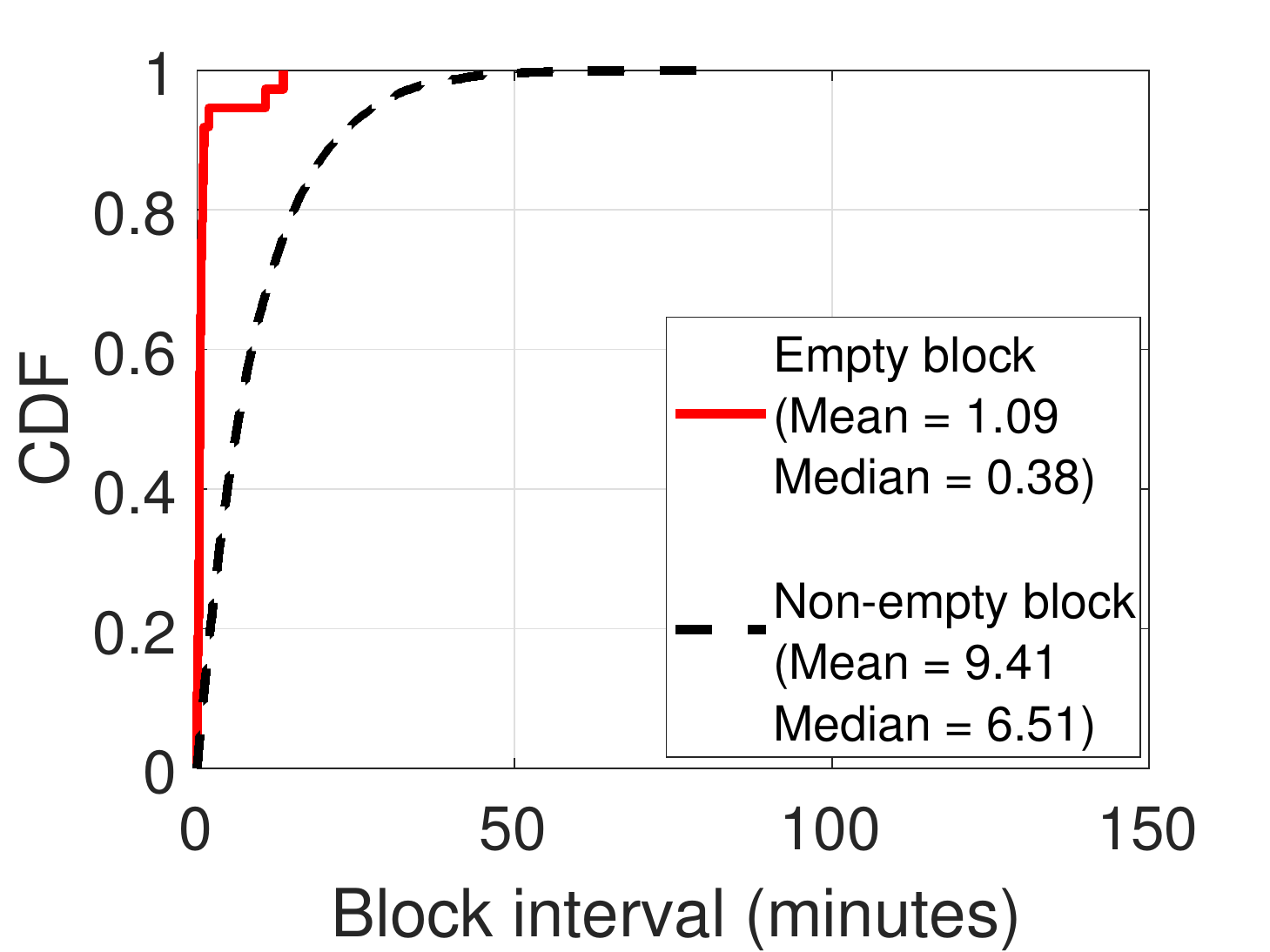}%
	}\hspace{0.2cm}
	\subfigure[BTC.com]{%
		\includegraphics[width=0.23\textwidth]{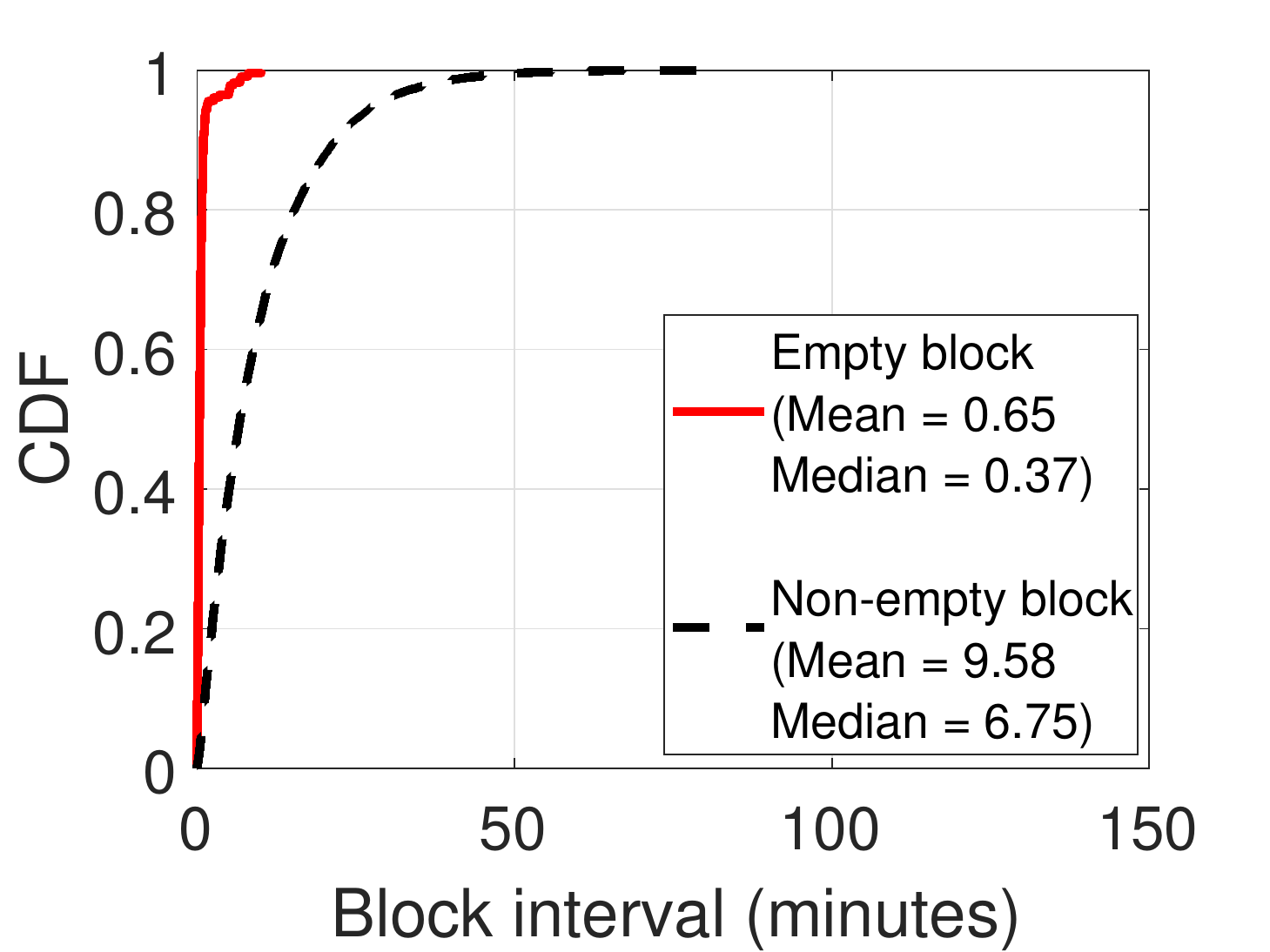}%
	}%
	
	\caption{Block Intervals of Empty Blocks and Non-Empty Blocks from Feb 25, 2016 to Jan 3, 2019}
\end{figure}

\textit{Reasons for Empty Blocks.} We are interested in reasons for empty blocks. As shown in TABLE 6 where empty block rates of different mining pools from February 25, 2016, to January 3, 2019, are presented. The first reason for the empty block is that the cost of mining an empty block is much less than the cost of mining a non-empty block since the block interval of an empty block is much shorter. Specifically, as shown in Fig. 10 where block intervals of empty blocks and non-empty blocks from Feb 25, 2016 to Jan 3, 2019 are presented. An interesting \textit{phenomenon} is that block intervals of empty blocks significantly less than block intervals of non-empty blocks. On average, block intervals of empty blocks are around 1 minute while block intervals of non-empty blocks take around 9.6 minutes on average. It indicates that generating empty blocks can significantly reduce block interval compared to generating non-empty blocks. Thus, following our model in section 4.3, the net profit of mining an empty block is given in equation (6) and the net profit of mining a non-empty block is in equation (7) as follows,

\begin{equation}
Net_{empty}(t_1,t_2)= \int_{t1}^{t2}{R_{empty}(t)-\upsilon \cdot H(t)}dt
\end{equation}

\begin{equation}
Net_{non-empty}(t_1,t_3)= \int_{t1}^{t3}{R_{non-empty}(t)-\upsilon \cdot H(t)}dt
\end{equation}

\noindent where the net profit of mining an empty block $Net_{empty}(t_1,t_2)$ is greater than the net profit of mining a non-empty block $Net_{non-empty}(t_1,t_3)$ when $R_{non-empty}(t)\approx R_{empty}(t)$ and $t_2-t_1\ll t_3-t_1$. The second reason is that for some mining pools, the progress of mining an empty block starts before getting the full block data as shown in Algorithm 3. This means mining an empty block can cut the time of receiving the full block data as well as the time of verifying all transactions of the block. More analysis based on block propagation time will be further discussed in section 5.1.

\textit{Empty Block Concerns.} Even though some people may argue that mining an empty blocks has no monetary advantages to those who mine non-empty blocks since empty blocks do not include transaction fees. Our measurement results show that the block intervals of empty blocks are significantly lower than the block intervals of non-empty blocks. It indicates that the cost of mining an empty block is significantly lower than the cost of mining a non-empty block. Especially when the transaction fees are so low that the net profit of an empty block is higher than the net profit of a non-empty block, empty blocks attract mining pools. However, high transaction fees are not attractive for the Bitcoin users. And empty blocks also harm users. Empty blocks result in relatively high transaction delay since empty blocks waste the transaction processing capability of the Bitcoin blockchain.

\section{Data Analysis on Unconfirmed Transactions}

\begin{algorithm}[t]
	
	{\footnotesize \SetKwInOut{Input}{Input}\SetKwInOut{Output}{Output}
		\Input{the latest \textit{block header} (required), the last \textit{full block} (optional).}
		\Output{New empty block(s) or non-empty block(s).}
		\BlankLine
		
		New Block \textit{Candidate$_1$}: start mining new empty blocks after receiving and verifying the last \textit{block header};\\
		
		New Block \textit{Candidate$_2$}: start mining new non-empty blocks after receiving and verifying the last \textit{full block}; \\
		
		\eIf{\rm Block time of \textit{Candidate$_1$} $<$ Block time of \textit{Candidate$_2$} }{
			generate empty block(s);
		}{
			generate non-empty block(s);
		}
	}
	
	\caption{{\footnotesize Empty Block Generating Algorithm}}
	
\end{algorithm}

Mining pools receive announcements of verified unconfirmed transactions from Bitcoin clients in the Bitcoin network, and get some transaction fees from these transactions to be included in the new block. In this section, we are interested in several properties of unconfirmed transactions such as the transaction delay and transaction collection strategy of mining pools.

\subsection{Transaction Delay}

\begin{figure}[t]
	\centering
	
	\subfigure[AntPool]{%
		\includegraphics[width=0.23\textwidth]{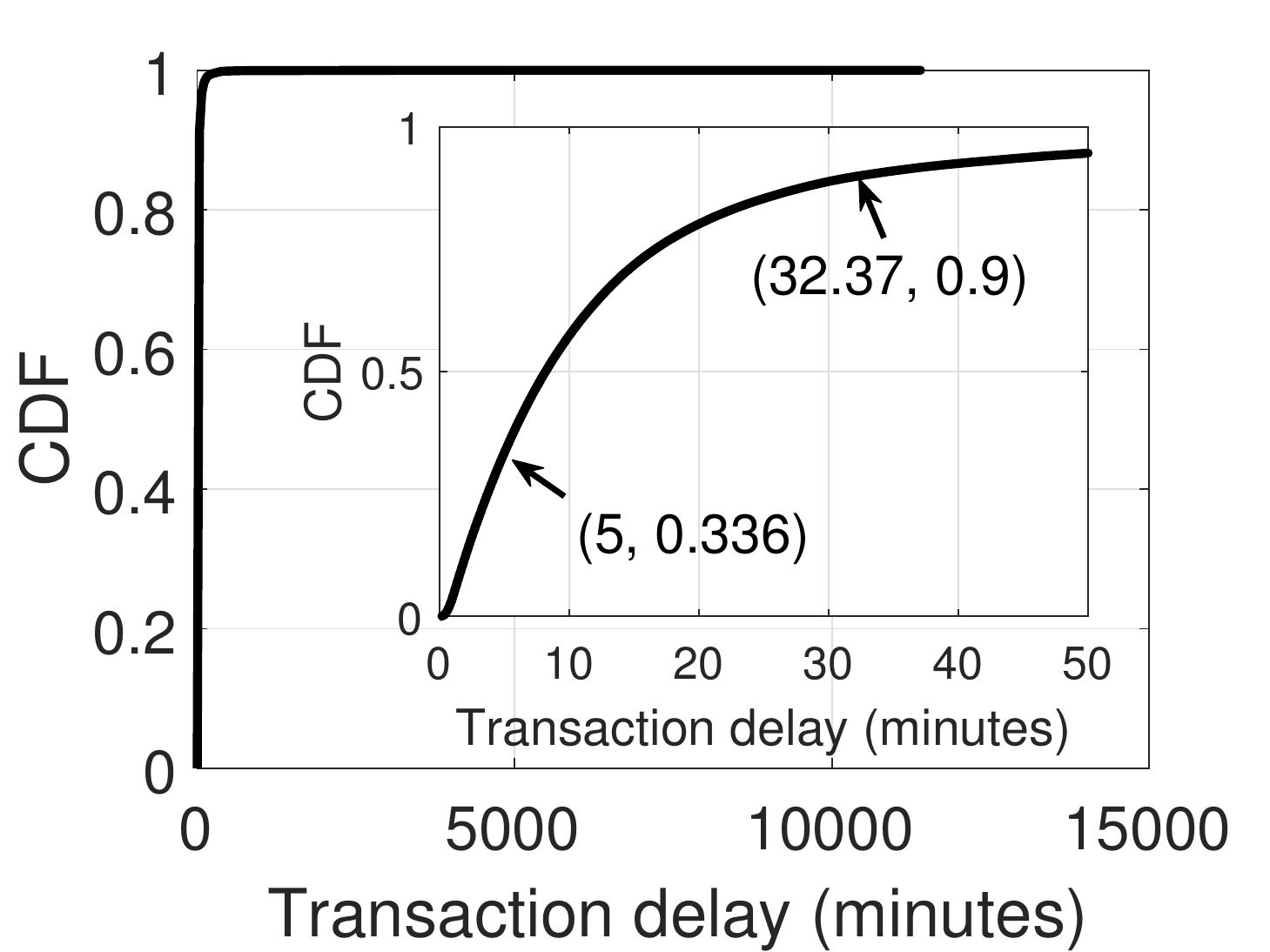}%
	}\hspace{0.2cm}
	\subfigure[F2Pool]{%
		\includegraphics[width=0.23\textwidth]{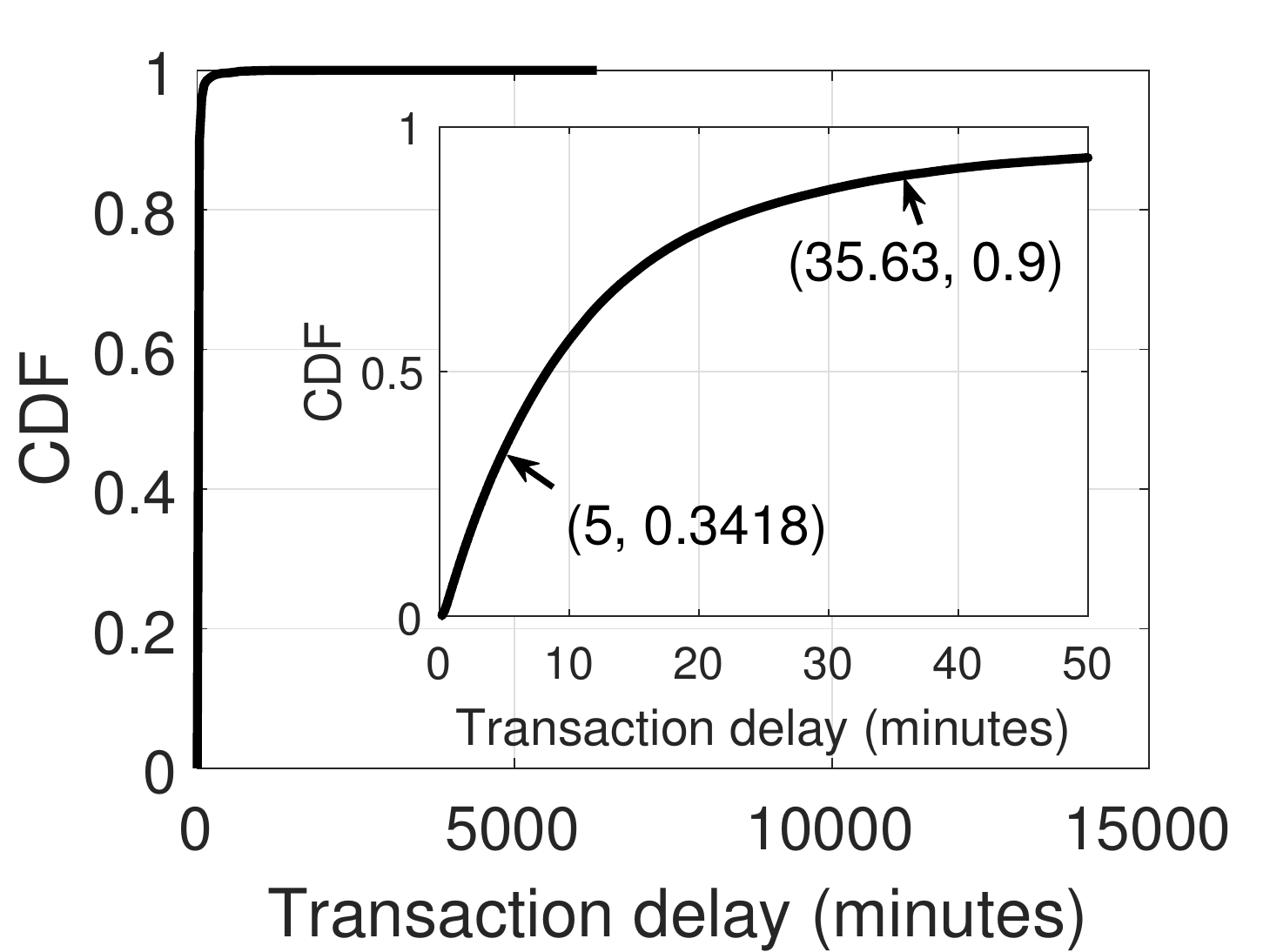}%
	}%
	
	
	\subfigure[ViaBTC]{%
		\includegraphics[width=0.23\textwidth]{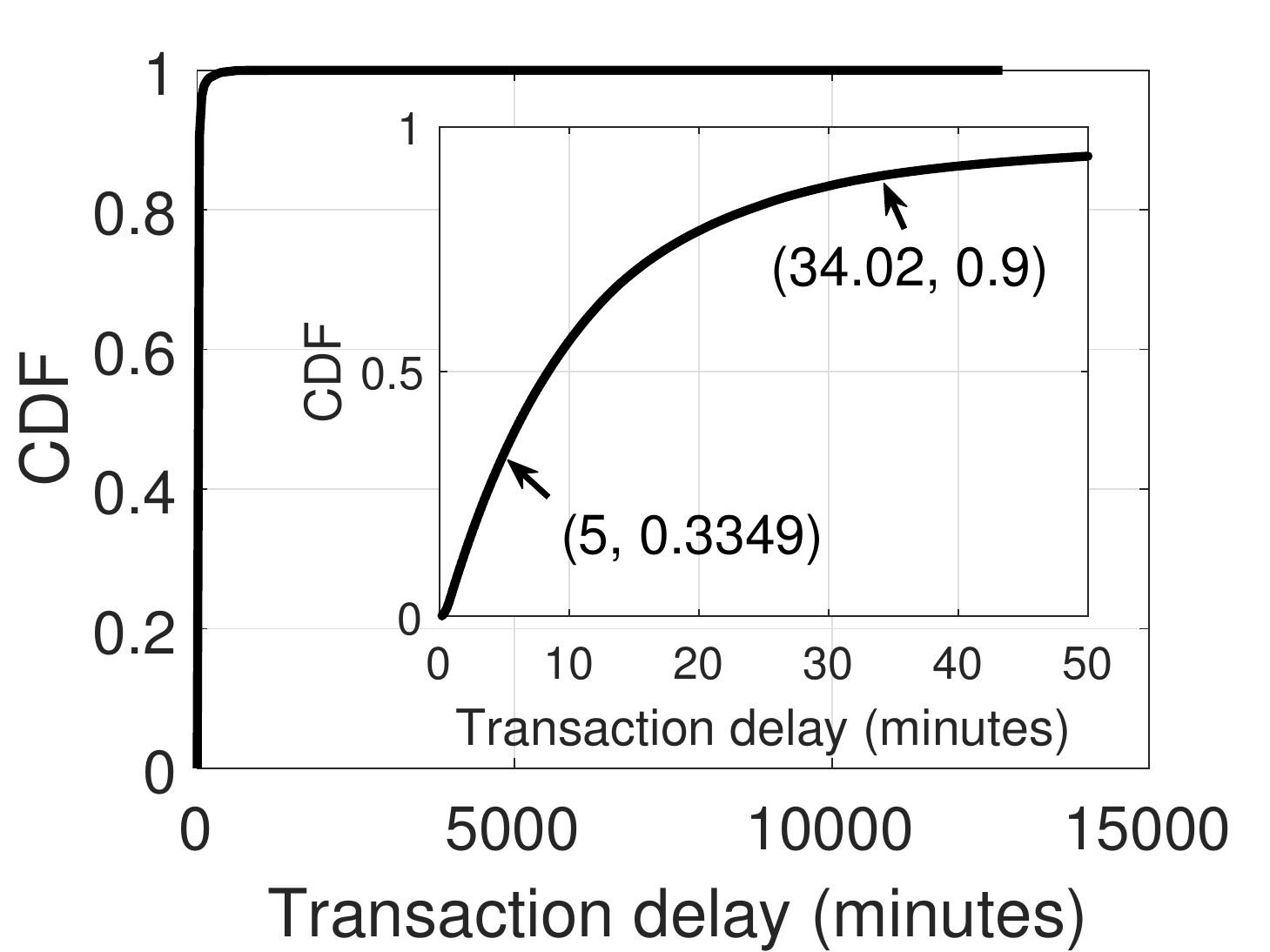}%
	}\hspace{0.2cm}
	\subfigure[BTC.com]{%
		\includegraphics[width=0.23\textwidth]{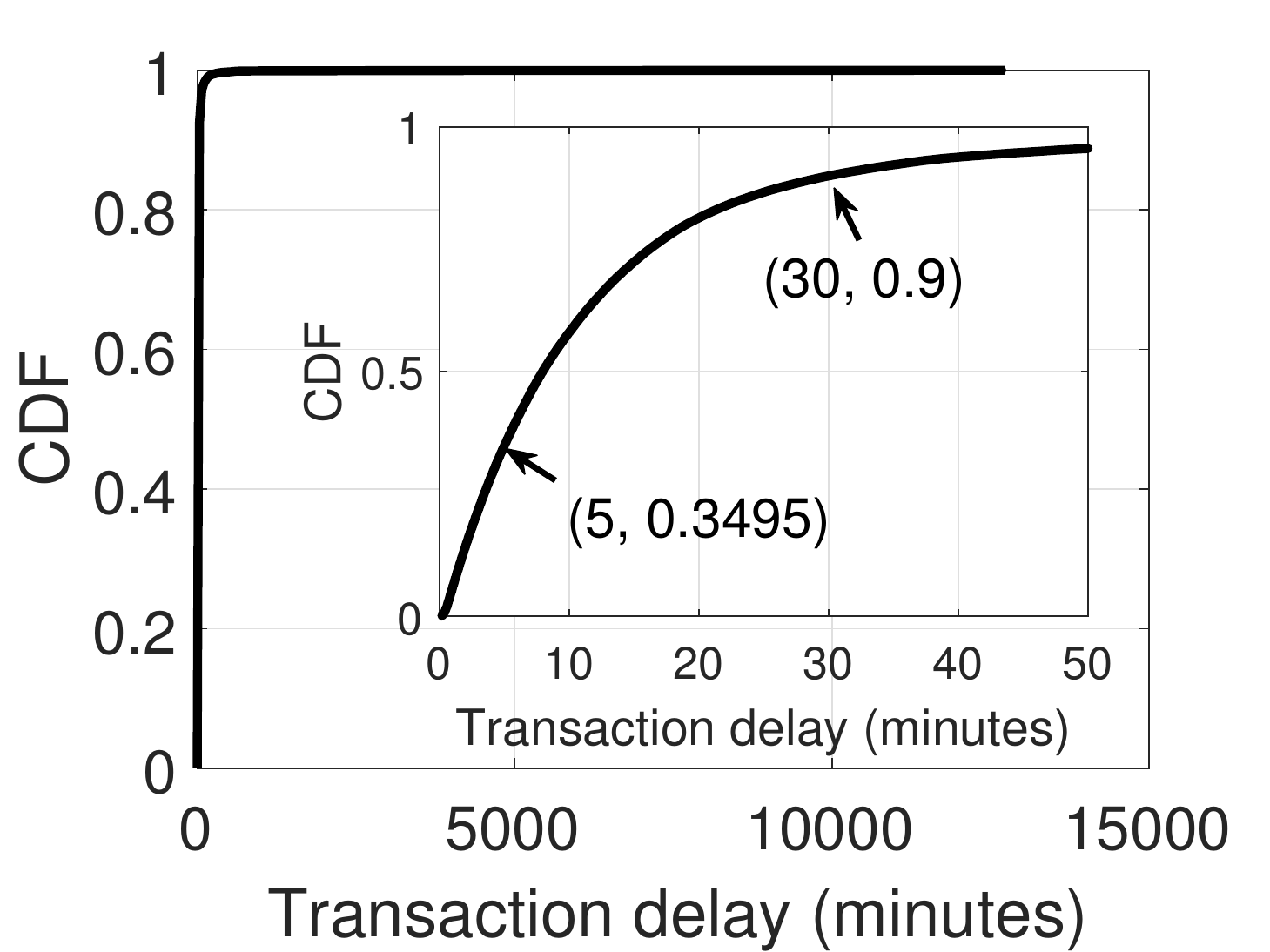}%
	}%
	
	
	\caption{Transaction Delay of Mining Pools from March 6, 2018 to January 3, 2019}
\end{figure}

Fig. 9 shows the process of broadcasting transactions and blocks data across the Bitcoin network. In step (a), peer A creates a new transaction named $tx_1$ and peer B creates a new transaction named $tx_2$. And these new transactions $tx_1$ and $tx_2$ are verified and stored at local Mempool, since they have not been confirmed yet. In step (b), peer A broadcasts the unconfirmed transaction $tx_1$ to neighbor peers B and C. Similarly, peer D broadcasts the $tx_2$ to neighbors peers B and C via the Bitcoin protocol. After transaction verification, peers B and C store these unconfirmed at local Mempool. In step (c), peer B collects unconfirmed transactions $tx_1$ and $tx_2$ from local Mempool, and include them into a new block named $b_n$. In step (d), peer B broadcasts the new block $b_n$ to neighbor peers A and C via the Bitcoin protocol. After block verification, peers A and C delete $tx_1$ and $tx_2$ from local Mempool since they have been confirmed by block $b_n$.

Transaction delay is the delay between the first time we observed an unconfirmed transaction, and the first time we observed that the transaction was written into a block. Bitcoin blocks are designed to be generated every ten minutes on average. The mining difficulty is dynamically adjusted depending on the time it took for solving previous blocks. Thus, the Bitcoin transaction delay should be around five minutes on average. In this part, we are interested in the transaction delay in the real world Bitcoin network.

Fig. 11 shows the transaction delay of mining pools from Mar 6, 2018 to Jan 3, 2019. We found that transaction delays vary widely. The longest transaction delays of mining pools could be a week or even longer during the observation period. But these transactions account for a very small proportion. For most of the transactions, the transaction delays are relatively shorter. For example, 90 percent of AntPool transactions delayed in 32.37 minutes and 33.6 percent of AntPool transactions delayed in 5 minutes. Other top mining pools such as F2Pool, ViaBTC and BTC.com also show the similar experimental results.

\subsection{Transaction Collection Strategy}

The transaction collection strategy is the strategy adopt by mining pools for collecting unconfirmed transactions into a new block. Several factors such as transaction size, transaction fee and waiting time, effect the transaction collection strategy. In this part, we are interested in the transaction collection strategy \cite{tschorsch2016bitcoin, pontiveros2018monitoring} for mining pools.

We proposed a simple model to analyze the the transaction collection strategy. Assuming mining pools are rational. The goal of the mining pool is to maximize its mining revenue. Thus, given a subset of $n$ unconfirmed transactions, each transaction $i$ has a transaction fee $p_i$ and a transaction size $w_i$. The maximum block size is $c$. In particular, $x_i=1$ means that mining pool chooses to collect transaction $i$; otherwise, $x_i=0$. Thus, we have the model as given in equation (8).  A solution to the model is the greedy algorithm. Mining pools pick the unconfirmed transactions with the largest \textit{feerate} (i.e., the ratio of transaction fee and transaction size), and it can get an approximate solution to the maximum mining profit.

\begin{figure}[t]
	\centering
	
	\subfigure[AntPool]{%
		\includegraphics[width=0.23\textwidth]{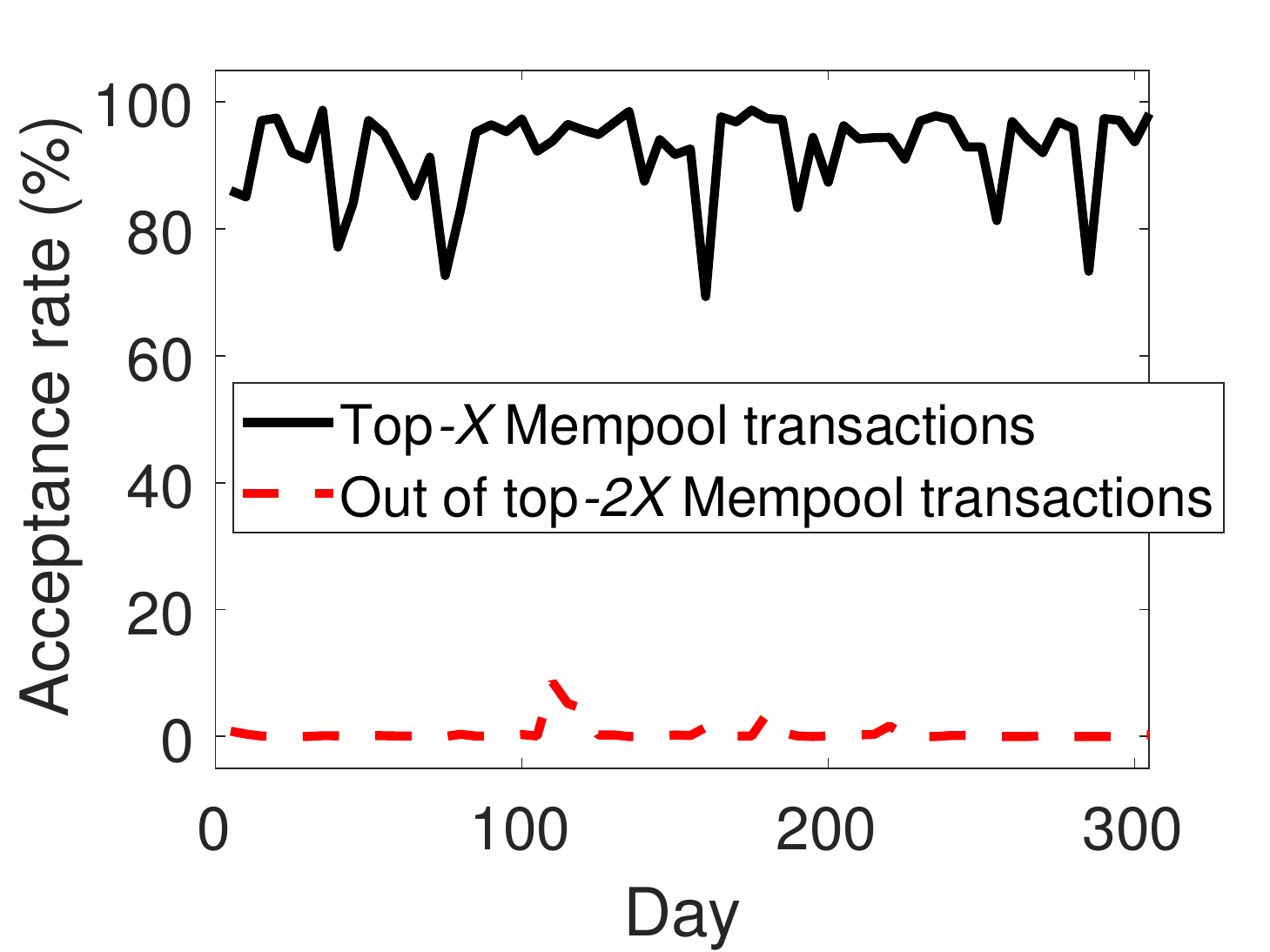}%
	}\hspace{0.2cm}
	\subfigure[F2Pool]{%
		\includegraphics[width=0.23\textwidth]{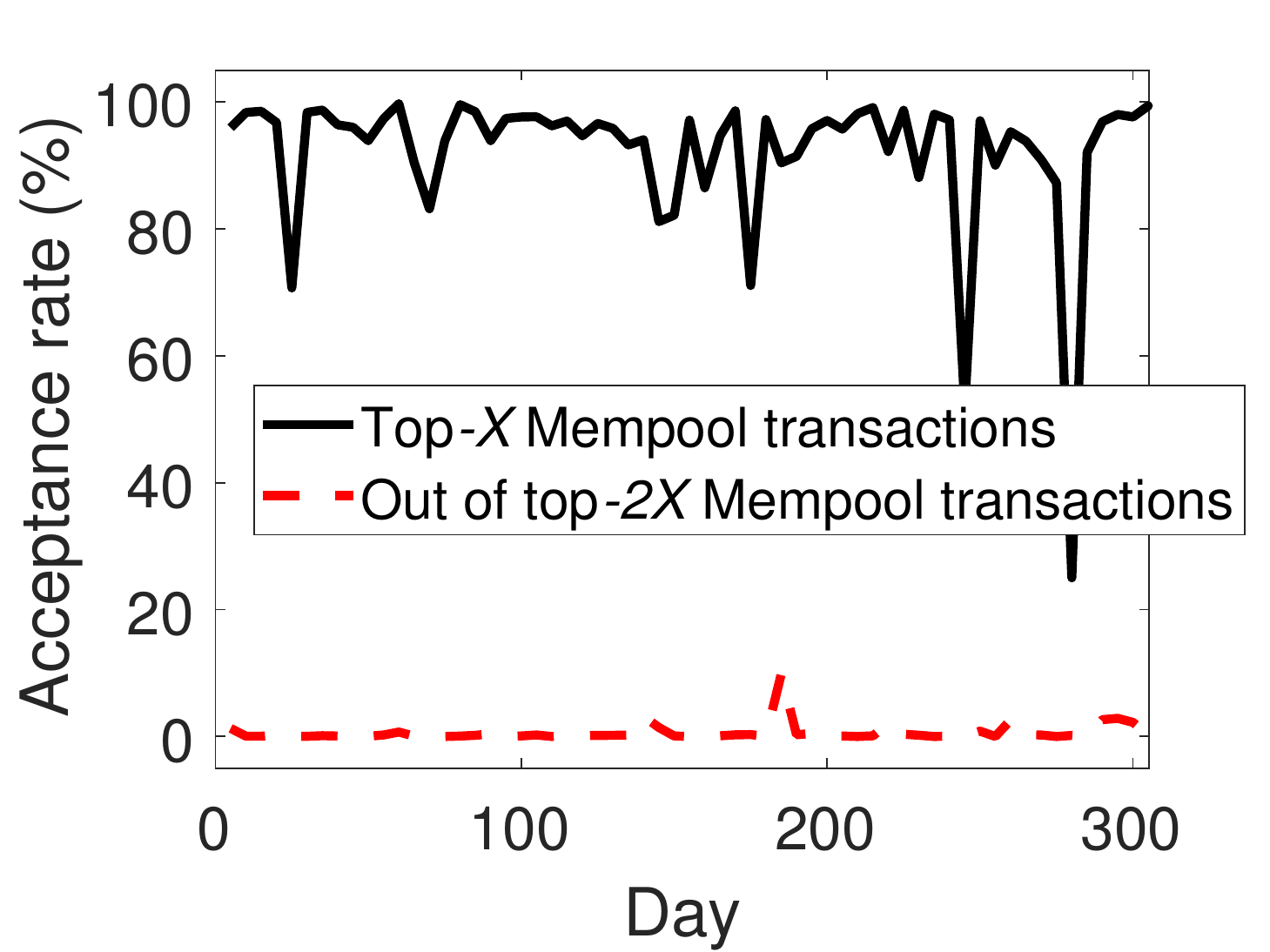}%
	}%
	
	
	\subfigure[ViaBTC]{%
		\includegraphics[width=0.23\textwidth]{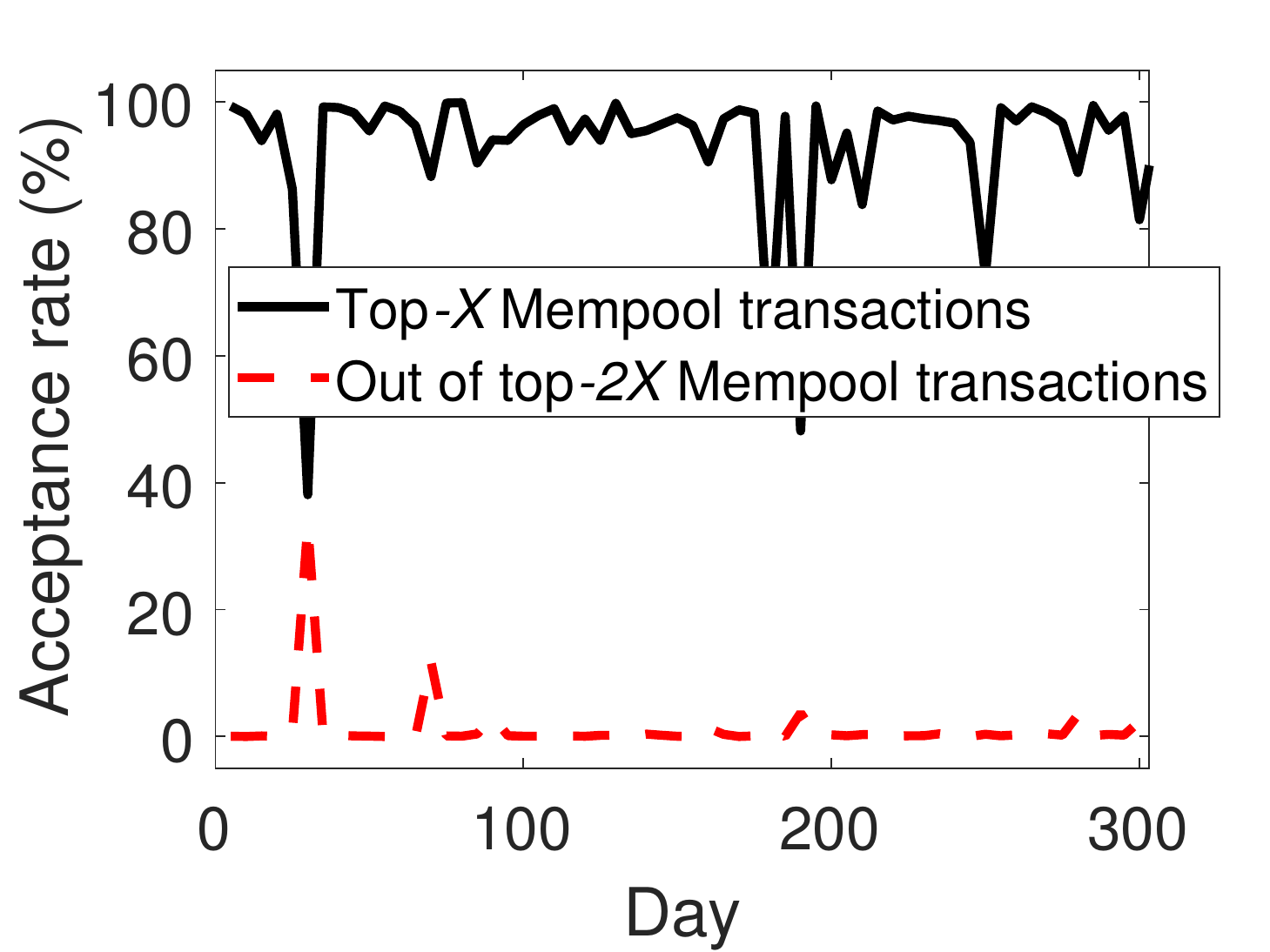}%
	}\hspace{0.2cm}
	\subfigure[BTC.com]{%
		\includegraphics[width=0.23\textwidth]{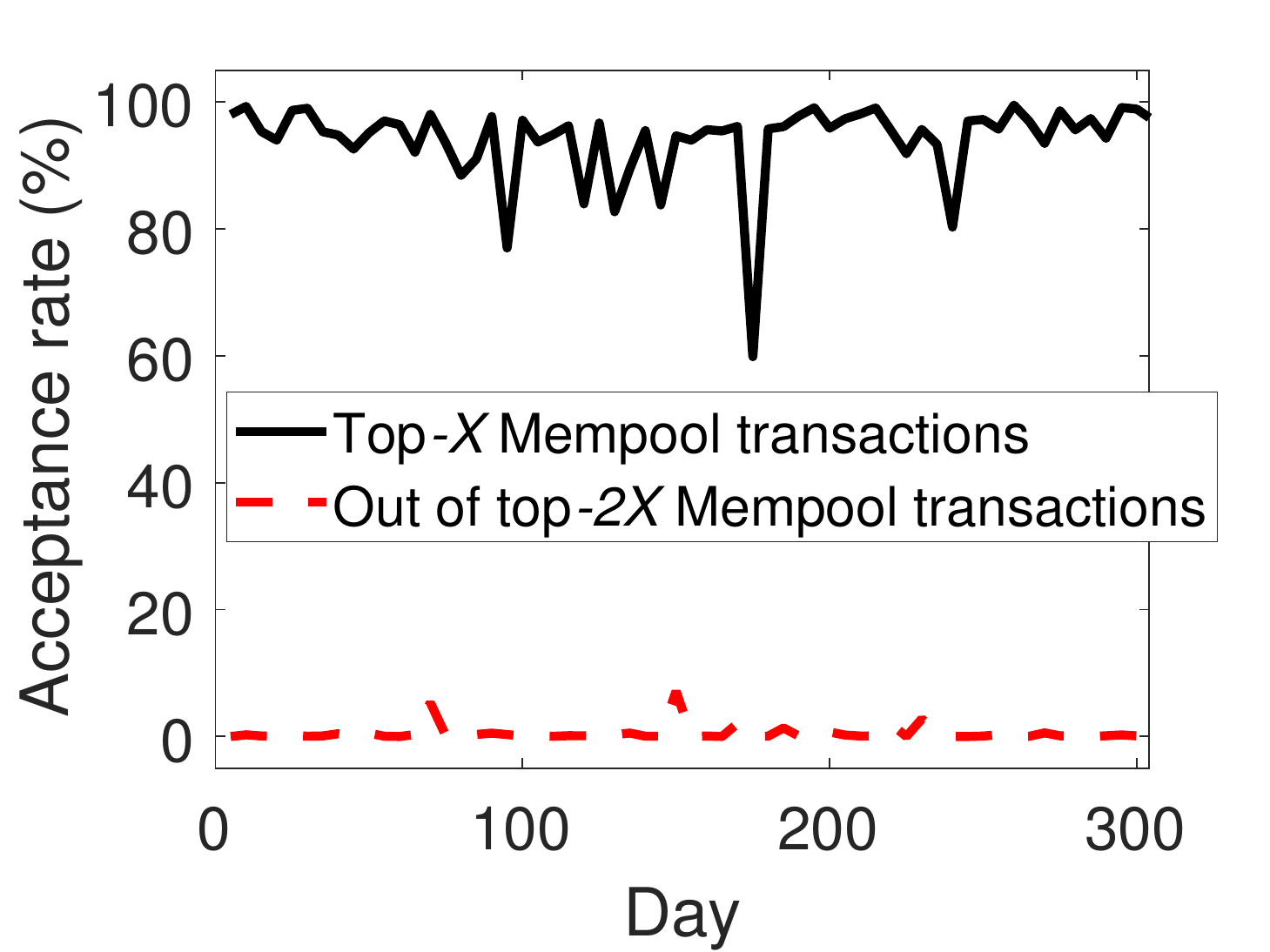}%
	}%
		
	\caption{Acceptance rate of local top-$X$ transactions with the highest \textit{feerate} from March 6, 2018 to January 3, 2019}
\end{figure}

\begin{equation}
\begin{array}{ll@{}ll}
\text{maximize}  & \displaystyle\sum\limits_{i=1}^{n} p_{i}x_{i} &\\
\text{subject to}& \displaystyle\sum_{i=1}^{n}w_ix_i\leq c&\\
&x_{i} \in \{0,1\}, &i \in \{1,2,3,...,n\}&
\end{array}
\end{equation}

\textit{Validation.} To detect the transaction collection strategy of mining pools and validate the importance of \textit{feerate} in transaction collection strategy, we developed a python tool for analyzing local Mempool unconfirmed transactions. Our methods are as follows. First, we sorted the local Mempool unconfirmed transactions according to their \textit{feerate} in a descending order. Second, we picked top-$X$ unconfirmed transactions from Mempool and constructed a data set of size $X$, where $X$ is the number of transactions of the next block. Third, we validated how many transactions of unconfirmed transactions are collected into the next block by mining pools and analyzed the properties of these unconfirmed transactions. Also, we introduced a new metric named the acceptance rate to describe the effectiveness of the \textit{feerate} strategy. For example, if the acceptance rate is 100 percent, it means 100 percent of the unconfirmed transactions are collected by mining pools. Otherwise, if the acceptance rate is 0 percent, it means none of the unconfirmed transactions are collected by mining pools. Through this approach, we can dissect the transaction collection strategy of different mining pools. 

Fig. 12 shows the acceptance rate of Mempool top-$X$ unconfirmed transactions with the highest \textit{feerate} from Mar 6, 2018 to Jan 3, 2019. We found that the \textit{feerate} is a dominating factor in the transaction collection strategy. Specifically, for AntPool, the acceptance rate of the unconfirmed transactions could be 89.46 percent on average if its \textit{feerate} ranks are in top $X$, On the converse, the acceptance rate of the unconfirmed transactions could be less than 3.81 percent if its \textit{feerate} ranks are out of top $2\times X$. Other mining pools such as F2Pool, ViaBTC and BTC.com also show the similar experimental results.

\section{Related Work}

\subsubsection{Bitcoin Network Measurements}

There are many existing works focus on the Bitcoin network. These measurement studies can be classified into two categories: network protocol analysis and network flow analysis. 

First, in terms of network protocol analysis, Gervais et al., \cite{gervais2015tampering} proposed that the adversary can exploit measurements in order to effectively delay the propagation of transactions and blocks to specific nodes for a considerable amount of time without causing a network partition. They also suggested some countermeasures in order to enhance the security of the Bitcoin network. Yonatan et al., \cite{sompolinsky2015secure} investigated the implications of having a higher transaction throughput on the Bitcoin security against double spending attacks. They also showed that at high throughput, substantially weaker attackers are able to reverse payments they made even well after they were considered accepted by recipients. Kiffer et al., \cite{kiffer2017stick} studied the large scale fork in the blockchain network and showed how the fork lead to unintentional incentives and security vulnerabilities. Toyoda et al., \cite{toyoda2017identification} identified Bitcoin addresses related with fraudulent activities such as high yield investment program by analyzing transactions patterns, and showed that about 83\% of fraudulent activities addresses are correctly classified while maintaining false positive rate less than 4.4\%. Croman et al., \cite{croman2016scaling} analyzed the performance bottlenecks of the Bitcoin protocol and suggested that parameters of block size and block intervals was the first increment toward achieving high local blockchain protocols. Chen et al., \cite{chen2017extending} presented a novel design statechain which used Bitcoin blockchain to propagate application log, enabling application nodes to efficiently query the log as well as tranfer log between blockchains. Luu et al., \cite{luu2017smart} proposed a novel protocol design for a decentralized mining pool, which incurs mining fees lower than centralized mining pools and is designed to scale to a large number of miners. Owenson et al., \cite{owenson2017proximity} introduced a proximity-aware extension to the Bitcoin protocol to improve the transaction propagation delay in the Bitcoin network by clustering nodes based on their membership. Neudecker et al., \cite{neudecker2015simulation} presented a simulation environment based on the Bitcoin network since computing power of the Bitcoin network are too huge to conduct real world experiments.

Second, in terms of network flow analysis, Joan et al., \cite{donet2014bitcoin} proposed a data collection process identifying more than 872,000 different Bitcoin nodes from which they presented the size of the Bitcoin network, the node geographic distribution and the network stability in terms of interrupted availability of nodes. Sallal et al., \cite{sallal2016bitcoin} measured transaction delay and analyzed transaction data propagation and considered data inconsistency of different nodes in the Bitcoin network. Qin et al., \cite{qin2018economic} provided a research framework to explore the economic issues in the Bitcoin ecosystems from the levels of mining pools, individual miners and blockchain network. Chen et al., \cite{chen2018understanding} conducted a measurement study on the Ethereum blockchain network by leveraging graph analysis to characterize three major activities namely money transfer, smart contract creation and smart contract invocation.

\subsubsection{Mining Pool Measurements}

An interesting topic in the Bitcoin network is mining pool measurements. Mining pools are major computing resources and significantly impact on both security and performance of the Bitcoin network. There are many measurement studies based on the Bitcoin mining pools, including network flow analysis \cite{wang2015exploring}, incentive mechanism and mining competition \cite{gervais2014bitcoin, lewenberg2015bitcoin, schrijvers2016incentive, zamyatin2017swimming, liu2018evolutionary}. Specifically, Wang et al., \cite{wang2015exploring} studied the Bitcoin network flows and presented the evolution of Bitcoin mining pools by conducting a case study on F2Pool. Arthur et al., \cite{gervais2014bitcoin} showed that vital operations and decisions that Bitcoin is currently undertaking are not decentralized, and showed that third-party entities can unilaterally decide to devalue any specific set of Bitcoin addresses pertaining to any entity participating in the system. Lewenberg et al., \cite{lewenberg2015bitcoin} focused on mining rewards distribution of the Bitcoin mining pools and showed that under high transaction loads, it is difficult or even impossible to distribute rewards in a stable way. Schrijvers et al., \cite{schrijvers2016incentive} introduced a game theoretic model for reward functions within a single Bitcoin mining pool and showed that PPLNS, a popular reward function, is incentive compatible in a more general model. Zamyatin et al., \cite{zamyatin2017swimming} formulated a model of the dynamics of a queue based reward distribution scheme. They showed that the underlying mechanism disadvantages miners with above-average hash rates. Liu et al., \cite{liu2018evolutionary} studied the dynamics of mining pool selection in a blockchain network, where mining pools may choose arbitrary block mining strategies. They demonstrated the stability in the evolution of miners' strategies in a general case.

\section{Conclusion}

This paper presented a thorough measurement study on the Bitcoin network. We traced over 1.56 hundred thousand blocks (including about 257 million historical transactions) from February 2016 to January 2019 and collected over 120.25 million unconfirmed transactions from March 2018 to January 2019. We then conducted an in-depth investigation of the Bitcoin network from a perspective of mining pools. We have shown that a few mining pools continuously control most of the computing resources of the Bitcoin network. We found that mining pools are caught in a prisoner's dilemma where mining pool competes to increase their computing resources even though the unit profit of the computing resource decreases. And mining pools are stuck in a Malthusian trap where there is a stage at which the Bitcoin incentives are inadequate for feeding the exponential growth of the computing resources. Moreover, we conducted a large scale measurement and analysis on Mempool unconfirmed transactions and found that \textit{feerate} plays a dominating role in transaction collection strategy for most of the top mining pools. In summary, the results of this study provide a comprehensive picture of the inside working of the Bitcoin network.

\ifCLASSOPTIONcaptionsoff
  \newpage
\fi



%
%
%

\bibliographystyle{IEEEtran}
\bibliography{tpds}

\newpage
\begin{IEEEbiography}[{\includegraphics[width=1in,height=1.25in,clip,keepaspectratio]{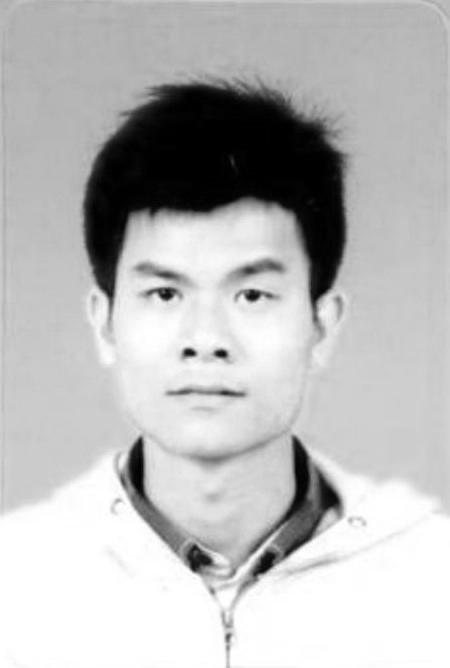}}]{Canhui Wang} received his B.E. degree from Shandong University, P.R. China, in 2016. He is currently pursuing his Ph.D. degree with Hong Kong Baptist University. He was a research assistant with IRACE research center in Shenzhen. His research interests include GPU computing, distributed storage systems and Blockchain technology. He is a graduate student member of the IEEE.
\end{IEEEbiography}
%
\vspace{-20pt}
\begin{IEEEbiography}[{\includegraphics[width=1in,height=1.25in,clip,keepaspectratio]{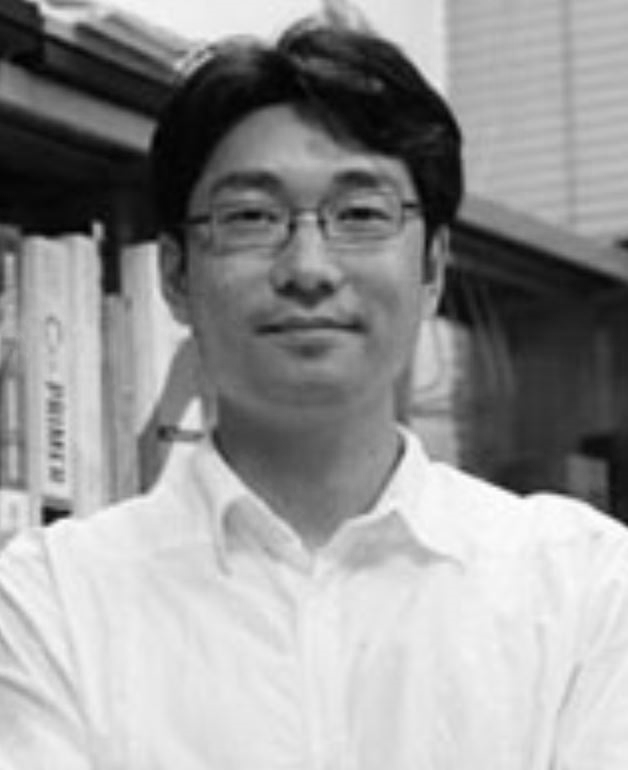}}]{Xiaowen Chu} received the B.E. degree in computer science from Tsinghua University, P.R. China, in 1999, and the Ph.D. degree in computer science from the Hong Kong University of Science and Technology in 2003. Currently, he is currently a full professor in the Department of Computer Science, Hong Kong Baptist University. His research interests include parallel and distributed computing, cloud computing and wireless networks. He is serving as an Associate Editor of IEEE Access and IEEE Internet of Things Journal. He is a senior member of the IEEE.
\end{IEEEbiography}
%
\vspace{-15pt}
\begin{IEEEbiography}[{\includegraphics[width=1in,height=1.25in,clip,keepaspectratio]{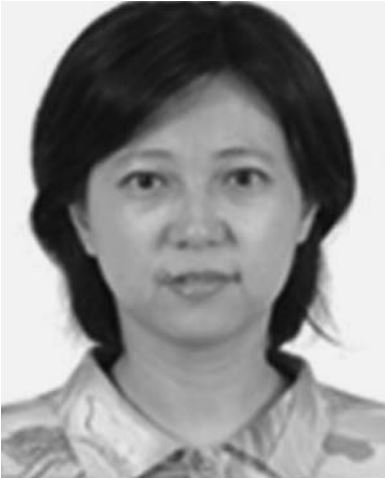}}]{Qin Yang} received her B.Sc in computer science at Southwest Jiaotong University, China, in 1989, M.S in computer science at Huazhong University of Science \& Technology, in 1992, and PhD in computer science, Hong Kong University of Science \& Technology, Kowloon, Hong Kong at November of 1999. She is an associate professor in the Department of Computer Science and Technology, Harbin Institute of Technology, Shenzhen Graduate School, China. Her research interest is in the area of wireless networks, mobile computing, crosslayer design, QoS of routing and scheduling, high speed optical networks and so forth. She is a senior member of the IEEE.
\end{IEEEbiography}

\vspace{240pt}

\end{document}